\documentclass[aps,prb,amsmath,amssymb,twocolumn,longbibliography,superscriptaddress,preprintnumbers]{revtex4-2}
\usepackage{float}
\usepackage[pdftex]{graphicx,xcolor}
\usepackage{dcolumn}
\usepackage{bm}
\usepackage{ascmac}
\usepackage{array}
\usepackage{siunitx}
\usepackage{physics}
\usepackage{hyperref}
\hypersetup{
     colorlinks=true,
     citecolor=blue,
     linkcolor=blue,
     }
\usepackage{times}

\usepackage[whole]{bxcjkjatype}

\newcommand{\s}{{\sigma}}

\newcommand{\bk}{{\bm{k}}}
\newcommand{\Z}{{\mathbb{Z}}}
\usepackage[normalem]{ulem} 

\usepackage{multirow}
\usepackage{bigdelim}

\begin{document}

\title {General corner charge formulas in various tetrahedral and cubic space groups}
\author {Hidetoshi Wada}
\affiliation{
Department of Physics, Tokyo Institute of Technology, 2-12-1 Ookayama, Meguro-ku, Tokyo 152-8551, Japan\\
}

\author {Katsuaki Naito}
\affiliation{
Department of Physics, Tokyo Institute of Technology, 2-12-1 Ookayama, Meguro-ku, Tokyo 152-8551, Japan\\
}

\author {Seishiro Ono}
\affiliation{
Interdisciplinary Theoretical and Mathematical Sciences Program (iTHEMS), Riken, Wako 351-0198, Japan\\
}

\author {Ken Shiozaki}
\affiliation{Center for Gravitational Physics and Quantum Information, Yukawa Institute for Theoretical Physics, Kyoto University, Kyoto 606-8502, Japan}

\author {Shuichi Murakami}
\affiliation{
Department of Physics, Tokyo Institute of Technology, 2-12-1 Ookayama, Meguro-ku, Tokyo 152-8551, Japan\\
}

\date{\today}
\preprint{YITP-23-108}
\preprint{RIKEN-iTHEMS-Report-23}

\begin{abstract}
     In some insulators, corner charges are fractionally quantized, due to the topological invariant called a filling anomaly. 
     The previous theories of fractional corner charges have been mostly limited to two-dimensional systems. 
     In three dimensions, only limited cases have been studied.
     In this study, we derive formulas for the filling anomaly and the corner charge in various crystals with all the tetrahedral and cubic space groups.
     We discuss that the quantized corner charge requires the crystal shapes to be vertex-transitive polyhedra.
     We show that the formula of the filling anomaly is universally given by the difference between electronic and ionic charges at the Wyckoff position $1a$.
     The fractional corner charges appear by equally distributing the filling anomaly to all the corners of the crystal.
     We also derive the $k$-space formulas for the fractional corner charge. 
     In some cases, the corner charge is not determined solely from the irreps at high-symmetry $k$-points. 
     In such cases, we introduce a new $\mathbb{Z}_{2}$ topological invariant to determine the corner charge.
\end{abstract}

\maketitle

\section{Introduction}
Topological phases of matter are characterized by gapless excitations localized at their boundaries and are robust against adiabatic deformations
\cite{PhysRevLett.98.106803,PhysRevLett.95.146802,PhysRevLett.95.226801,PhysRevB.76.045302,PhysRevB.74.195312,PhysRevB.27.6083,doi:10.1126/science.1148047,doi:10.1126/science.1133734,PhysRevB.78.045426,PhysRevLett.106.106802,PhysRevB.96.245115,PhysRevB.91.161105,PhysRevB.95.081107,PhysRevB.90.165114,PhysRevB.98.081110,PhysRevLett.119.246402,PhysRevResearch.2.043131,PhysRevB.105.045126,PhysRevB.99.245151,Schindler_2018,doi:10.1126/sciadv.aat0346,PhysRevResearch.2.012067,PhysRevLett.124.036803,PhysRevX.11.041064,PhysRevB.103.205123,PhysRevResearch.1.033074,PhysRevB.102.165120,10.1093/acprof:oso/9780199564842.001.0001,PhysRevB.78.195125,PhysRevLett.123.066404,Zhang_2022,PhysRevB.107.195112,Bradlyn_2017,PhysRevB.97.035139,Aroyo:xo5013,Po_2017}. 
Today, they form an important foundation of condensed matter physics, and research on topological insulators has been attracting a lot of attention. 
In recent years, varieties of topological insulators have emerged, such as topological crystalline insulators (TCIs) protected by crystallographic symmetries and higher-order topological insulators (HOTIs) with gapless states on their hinges or corners~\cite{PhysRevB.78.045426,Slager:2013aa,PhysRevLett.106.106802,PhysRevB.96.245115,PhysRevB.91.161105,PhysRevB.95.081107,PhysRevB.90.165114,PhysRevB.98.081110,PhysRevLett.119.246402,PhysRevResearch.2.043131,PhysRevX.7.041069,PhysRevB.105.045126,PhysRevB.99.245151, PhysRevB.98.081110,PhysRevLett.119.246402,PhysRevResearch.2.043131,PhysRevB.105.045126,PhysRevB.99.245151,Schindler_2018,doi:10.1126/sciadv.aat0346,PhysRevResearch.2.012067,PhysRevLett.124.036803,PhysRevX.11.041064,PhysRevB.103.205123,PhysRevResearch.1.033074,PhysRevB.102.165120}.
Furthermore, topological quantum chemistry \cite{Bradlyn_2017,PhysRevB.97.035139,Aroyo:xo5013, MTQC} and theory of symmetry-based indicators \cite{Po_2017, SI_SA_Watanabe} have enabled us to identify many topological materials~\cite{catalogue0,catalogue1,catalogue2,catalogue3,catalogue4,catalogue6}.

Unlike topological insulators, topologically trivial insulators called atomic insulators (AIs) have been believed to have no topological features because their energy spectra are completely gapped including hinges and corners. 
However, recent studies have revealed that some AIs have fractionally quantized charges on their corners \cite{PhysRevB.105.045126,PhysRevB.99.245151,PhysRevX.11.041064,PhysRevB.103.205123,PhysRevResearch.1.033074,PhysRevB.102.165120}. 
One example of those AIs is an obstructed atomic insulator (OAI).
Electrons in AIs occupy exponentially localized Wannier orbitals, and OAIs are characterized by charge imbalance associated with each Wyckoff position. 
Fractional corner charges in OAIs can be understood in terms of a filling anomaly \cite{PhysRevB.99.245151,PhysRevB.103.165109}. 
A filling anomaly is a topological invariant given by the difference between the number of electrons required by symmetries of the system such as rotational symmetry of the system and that required by charge neutrality. 
Fractional corner charges can be explained by the equivalent distribution of the filling anomaly to the corners.

In the previous works \cite{PhysRevB.105.045126}, some of the authors derived formulas of the hinge charge and the corner charge formulas for the space group (SG) No.~207 for five crystal shapes of vertex-transitive polyhedra such as a cube, an octahedron, and a cuboctahedron with cubic symmetry in terms of both bulk Wyckoff positions and bulk band structures. 
Compared to the corner charge formula for two-dimensional systems, that for three-dimensional systems is complex in the sense that there are many varieties of crystal shapes for the same point-group symmetry in three dimensions, and we must consider the charge neutrality conditions not only of the bulk and the surfaces but also of the hinges. 
Therefore we assumed perfect crystals with the point-group symmetry $O$, and determined the formula of the filling anomaly by counting the total numbers of electrons and ions given the positions of ions and Wannier centers of the electrons. 
From the filling anomaly, we could extract formulas of the hinge charge and corner charge. 
Moreover, to obtain the corner charge formulas from bulk band structures, we took the method of the elementary band representation (EBR) matrix, which gives the relationship between the irreducible representations (irreps) at Wyckoff positions in real space and at the high symmetry points (HSPs) in a momentum space \cite{PhysRevB.103.165109}. 
In this case, we found that the corner charge formulas cannot be determined only from the band structures, because some OAIs with different Wyckoff positions have the same irreps at all the HSPs in the momentum space. 
Therefore we introduced a Wilson-loop invariant $\Xi$, and successfully derived the corner charge formulas by incorporating it into EBR matrix. 
However, it is difficult to find this invariant for other SGs, and the corner charge formulas were not perfect.  

In this study, we focus on all SGs with cubic symmetry, No.~195 -- 230. 
Here, the SG numbers 198, 199, 205, 206, 212, 213, 214, 220, and 230 are excluded, because not all the corners of these nine SGs are not equivalent and therefore the corner charge is not quantized. 
We find that in all the vertex-transitive polyhedra corresponding to each SG, the corner charge formula is universally given by the filling anomaly divided by the number of corners of crystals with the above SGs.  
We also derive the corner charge formulas of those SGs in terms of the $k$-space wave functions by utilizing the $\Z_2$ topological invariant introduced in Ref.\cite{Ono=Shiozaki_top_inv}. 
This topological invariant is derived for the SG No.~16 with time-reversal symmetry, which is a subgroup of those SGs. 
The role of time-reversal symmetry\ (TRS) is discussed in Sec.~\ref{sec:Corner charge formulas using a new topological invariant}.

This paper is organized as follows. 
In Sec.~\ref{Sec.2}, we review the previous work by some of the authors on fractional hinge and corner charges in various crystal shapes. 
In Sec.~\ref{Sec.3}, we construct our theory on the filling anomaly in all the cubic SGs under various crystal shapes and derive formulas for hinge and corner charges. 
In Sec.~\ref{Sec.4}, we construct a $k$-space formula by means of the EBR matrix and a $\mathbb{Z}_{2}$ topological invariant. 
Our conclusion is given in Sec.~\ref{Sec.5}.

\section{Review of the fractional hinge and corner charges in various crystal shapes with the SG No.~207 \label{Sec.2}}
\begin{figure}
    \centering
    \includegraphics[scale = 0.76]{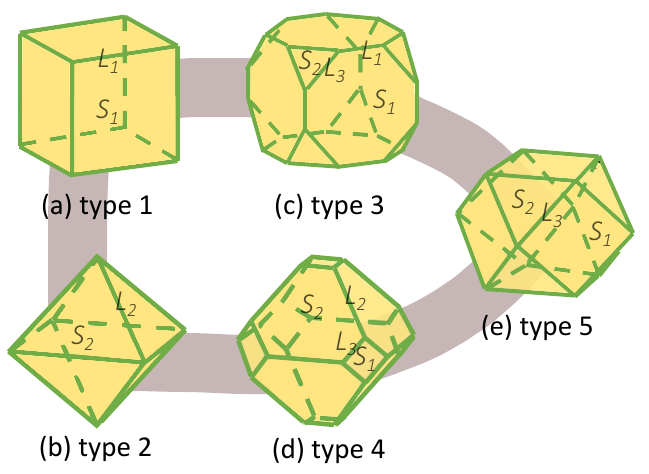}
     \caption{
     Five crystal shapes considered in the previous work~\cite{PhysRevB.105.045126}, all of which are vertex-transitive polyhedra. Their centers are placed at the Wyckoff position $1a$. (a) type 1: a cube. (b) type 2: a regular octahedron. (c) type 3: a truncated cube. (d) type 4: a truncated octahedron. (e) type 5: a cuboctahedron. The surfaces $S_{1}$ and $S_{2}$ have the Miller index $\{100\}$ and $\{111\}$, respectively. The hinge $L_{1}$ is an intersection of two $\{100\}$ surfaces. The hinge $L_{2}$ is an intersection of two $\{111\}$ surfaces. The hinge $L_{3}$ is an intersection of a $\{100\}$ surface and a $\{111\}$ surface. The purple line represents the path of continuous deformation of the vertex-transitive polyhedra.}
     \label{fig:crystal_shapes}
\end{figure}

In this section, we briefly review the previous work \cite{PhysRevB.105.045126} by some of the authors on fractional corner charges for the five crystal shapes shown in Fig.~\ref{fig:crystal_shapes} under SG $P432$ (No.~207). 
This SG is generated by fourfold rotation along $z$-axis, threefold rotation along $(111)$-direction, and translations.
In this section, we assume that the ground state of insulators is an atomic insulator whose occupied states are described in terms of exponentially localized Wannier states.
We explain how to derive the filling anomaly directly in Ref.~\cite{PhysRevB.105.045126}, and obtain the corner charge formulas.

We note that the filling anomaly and corner charge formulas in terms of bulk Wyckoff positions described below are also applicable to the ground state of fragile topological insulators~\cite{FragileTI_Po_Watanabe_Vishwanath}.
A fragile topological insulator is an insulator that becomes an atomic insulator when directly summed with some atomic insulator but itself has no exponentially localized Wannier representations. 
Nevertheless, as we explain in Sec.~\ref{IVA}, as far as the filling anomaly is concerned, a fragile topological insulator can be treated as an atomic insulator.

\subsection{Preparations for the derivation of corner charge formulas}
\begin{figure}
    \centering
    \includegraphics[scale = 0.8]{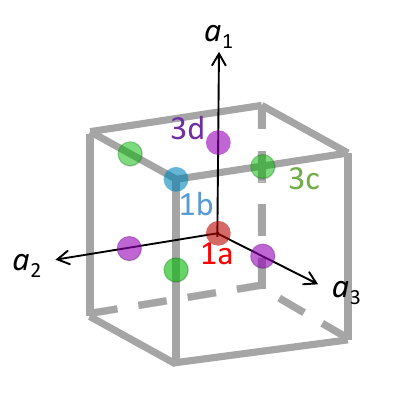}
    \caption{Wyckoff positions in the space group No.~207 in the previous work~\cite{PhysRevB.105.045126}. The points with the same colors belong to the same Wyckoff positions. $1a$, $1b$, one of $3c$ and one of $3d$ are positioned at $(0,0,0)$, $\frac{\bm{a}_{1}+\bm{a}_{2}+\bm{a}_{3}}{2}$, $\frac{\bm{a}_{1}+\bm{a}_{2}}{2}$ and $\frac{\bm{a}_{1}}{2}$, respectively, where $\bm{a}_{i}\ (i = 1, 2, 3)$ is  primitive lattice vectors in the cubic unit cells. The gray cube represents the unit cell.}
    \label{fig:unit_cell}
\end{figure}
Before deriving corner charge formulas for the five crystal shapes with SG No.~207, we discuss the properties of electric charges in AIs. 
Electric charges in AIs consist of ions and electrons, and electrons occupy Wannier orbitals of electronic bands.
Because ions are made of nuclei and core electrons, the ionic charges in AIs are integer multiples of the elementary charge $\abs{e}$, where $-e\ (e>0)$ is the electron charge.
Moreover, in AIs, Wannier orbitals are exponentially localized, and the integral charges of electrons in the Wannier orbitals can be assigned to the Wannier centers, where the gap is open both in the bulk and the boundaries. 

We also review the filling anomaly. 
In finite-sized crystals of certain insulators, there exist cases where the charge neutrality condition is inconsistent with crystallographic symmetry. 
In this case, by adding or removing some electrons from charge neutrality, the entire system is insulating while maintaining the symmetry.
The number of electrons required for this operation is called the filling anomaly. 

Now, we discuss the filling anomaly for cubic crystals. 
Let $n_{\omega}$ be the number of Wannier orbitals at a maximal Wyckoff position $\omega (= a, b, c, d)$ in the unit cell of the lattice as shown in Fig.~\ref{fig:unit_cell} and $m_{\omega}\abs{e}$ be the total ionic charge at a maximal Wyckoff position $\omega$.
Then, an integer $\Delta\omega$ is defined by the difference between them, i.e., 
\begin{align}
    \Delta\omega = n_{\omega} - m_{\omega} \in \mathbb{Z}.
    \label{eq:delta}
\end{align}
We note that other Wyckoff positions, except for the maximal ones, can be deformed to the maximal ones continuously.

Thus when calculating the filling anomaly, all we need to do is count the number of electrons at the Wannier centers and the number of ions at each maximal Wyckoff position in the cubic unit cell as shown in Fig.~\ref{fig:unit_cell}.
To obtain the filling anomaly, we take the center of a crystal to be the maximal Wyckoff position $1a$, and assume a perfect crystal, which implies that there is no surface reconstruction so that the periodicities of the hinge and the surface reflect that of the bulk. 
Then, the filling anomaly $\eta_{n}$ for a finite-sized crystal with $n$ hinge unit cells can be expressed as a polynomial for $n$:
\begin{align}
    \eta_{n} = \alpha_{3} n^{3} + \alpha_{2} n^{2} + \alpha_{1} n + \alpha_{0}, 
    \label{eq:filling_anomaly}
\end{align}
where $\alpha_{i} (i = 0, 1, 2, 3)$ are constant coefficients that depend on crystal shapes.

\subsection{Corner charge formulas in terms of the bulk Wyckoff positions}
Here for simplicity, we will only consider the crystal with the shape of a cube (type 1 in Fig.~\ref{fig:crystal_shapes}) with $n$ unit cells along each hinge. Using Eqs.~(\ref{eq:delta}) and (\ref{eq:filling_anomaly}), the filling anomaly is given by
\begin{equation}
    \begin{aligned}[b]
        \eta^{\text{perfect,type 1}}_{n} = 
        & (\Delta a + \Delta b + 3\Delta c + 3\Delta d) n^{3} \\
        & + 3(\Delta b + \Delta d + 2\Delta c) n^{2} \\
        & + 3(\Delta b + \Delta c) n + \Delta b.
        \label{eq:cubic_filling_anomaly}
    \end{aligned}
\end{equation}
From the first term on the right-hand side, the bulk charge density $\rho_{\text{bulk}}$ (per bulk unit cell) 
is given by $ \rho_{\text{bulk}} = -\abs{e}(\Delta a + \Delta b + 3\Delta c + 3\Delta d)$.
Here we impose the charge neutrality condition for the bulk, given by
\begin{align}
    \rho_{\text{bulk}} = -\abs{e}(\Delta a + \Delta b + 3\Delta c + 3\Delta d) = 0.
    \label{eq:bulk_charge}
\end{align}
Next, since the surface area consists of $6n^2$ surface unit cells, 
the second term of Eq.~(\ref{eq:cubic_filling_anomaly}) gives 
the surface charge density $\sigma_{\text{sur}}$ (per surface unit cell) to be $ \sigma_{\text{sur}} \equiv -\abs{e}\frac{3(\Delta b + \Delta d + 2\Delta c)}{6}\ (\text{mod}\  \abs{e} )$. 
The surface charge density $\sigma_{\text{sur}}$ is proportional to the bulk polarization, and therefore, based on the modern theory of polarization \cite{PhysRevB.48.4442,PhysRevB.47.1651}, the ambiguities of the surface charge density per surface unit cell is given in terms of modulo $\abs{e}$. 
Now, in order to define the corner charge, we impose the charge neutrality condition for the surface of the type 1 crystal by 
\begin{equation}
    \begin{aligned}[b]
        \sigma_{\text{sur}} &\equiv -\abs{e}\frac{3(\Delta b + \Delta d + 2\Delta c)}{6} \\ &\equiv -\abs{e}  \frac{\Delta b + \Delta d}{2} = 0 \ (\text{mod}\  \abs{e} ). \label{eq:surface_charge}
    \end{aligned}
\end{equation}

Thus, when we assume the charge neutrality for bulk and surfaces, the first and second terms on the right-hand side in Eq.~(\ref{eq:cubic_filling_anomaly}) can be dropped:
\begin{align}
    \eta^{\text{perfect,type 1}}_{n} = 3(\Delta b + \Delta c) n + \Delta b.
    \label{eq:modified_filling_anomaly}
\end{align}
The first term on the right-hand side represents the hinge part.
Since the hinges consist of 
$12n$ unit cells along the hinge, 
we obtain the hinge charge density $\lambda^{L_{1}}_{\text{hinge}}$ per unit cell along the hinge:
\begin{equation}
    \begin{aligned}[b]
        \lambda^{L_{1}}_{\text{hinge}} &= - \frac{\Delta b + \Delta c}{4} \abs{e} \\
        &\equiv - \frac{\Delta a + \Delta d}{4} \abs{e}\ (\text{mod}\  \abs{e}).
        \label{eq:hinge_charge}
    \end{aligned}
\end{equation}
To obtain the quantized corner charges, we need to introduce the charge neutrality condition for the hinge:
\begin{align}
    \Delta a + \Delta d \equiv 0 \ (\text{mod}\  4).
    \label{eq:hinge_charge_2}
\end{align}
Therefore, we find the corner charge $Q^{\text{type 1}}_{\text{corner}}$:
\begin{align}
    Q^{\text{type 1}}_{\text{corner}} = - \frac{\Delta b}{8} \abs{e} = - \frac{\Delta a}{8} \abs{e} \ (\text{mod}\  \frac{\abs{e}}{4}).
\end{align}
Here, we used the relation $\Delta a \equiv \Delta b \equiv \Delta c \equiv \Delta d \ (\text{mod}\ 2)$ under Eqs.~\eqref{eq:bulk_charge}, \eqref{eq:surface_charge}, and \eqref{eq:hinge_charge_2}. 

By doing the same calculations for the type 2 crystal (octahedron), we obtain 
\begin{align}
    \lambda^{L_{2}}_{\text{hinge}} = \frac{\Delta a + \Delta d}{2} \abs{e}\ (\text{mod}\ \frac{\abs{e}}{3}), \\
    Q^{\text{type 2}}_{\text{corner}} =  - \frac{\Delta a}{6} \abs{e} \ (\text{mod}\  \frac{\abs{e}}{3}).
\end{align}

Based on the results obtained from type 1 and type 2 crystals, we can get the hinge charge density $\lambda^{L_{3}}_{\text{hinge}}$ and the corner charges for the crystal shapes of the types 3--5 in Fig.~\ref{fig:crystal_shapes}:
\begin{align}
    \lambda^{L_{3}}_{\text{hinge}} = -\frac{\Delta a + \Delta d}{4} \abs{e}\ (\text{mod}\ \frac{\abs{e}}{6}), \\
    Q^{\text{type 3}}_{\text{corner}} =  - \frac{\Delta a}{24} \abs{e} \ (\text{mod}\  \frac{\abs{e}}{12}), \\
    Q^{\text{type 4}}_{\text{corner}} =  - \frac{\Delta a}{24} \abs{e} \ (\text{mod}\  \frac{\abs{e}}{12}), \\
    Q^{\text{type 5}}_{\text{corner}} =  - \frac{\Delta a}{12} \abs{e} \ (\text{mod}\  \frac{\abs{e}}{6}).
\end{align}
Thus, for the crystal shapes of the types 1--5, the corner charge formula with SG No.~207 in terms of the bulk Wyckoff positions is summarized
\begin{align}
    Q_{\text{corner}} = - \frac{\Delta a}{N}\abs{e}\ (\text{mod}\ \frac{2\abs{e}}{N}),
\end{align}
where $N$ represents the number of the corners of the crystal shapes of the types 1--5.

\section{Extension to other crystal shapes with other crystalline symmetries \label{Sec.3}}
\begin{table*}[t]
\caption{Summary for the point groups at the Wyckoff position $a$ and the charge densities of surfaces, hinges, and corners for the 27 space groups. For all the space groups we consider, surface charge densities are independent of the Miller indices of the surfaces appearing in a cube, an octahedron, and a tetrahedron. Because all the hinge charge densities of octahedrons are equal to those of tetrahedrons for every space group, such charge density is represented together in the column $\lambda_{\mbox{\scriptsize hinge}}^{L_{2}}$. The number of the corners is represented by $N$. We note that our corner charge formulas are only valid for the crystal shapes connected by any symmetry operations corresponding to the point group at the Wyckoff position $a$ shown in Fig.~\ref{fig:polyhedra}. We note that the value of $Q_{\mbox{corner}}$ is universally given by $\frac{\Delta a}{N}$ for all the cases.
}
\centering
\renewcommand{\arraystretch}{1.2}
\begin{ruledtabular}
\begin{tabular}{cccccc}
 SG number & the point group of Wyckoff position $a$ & $\sigma_{\mbox{sur}}$ (mod 1) & $\lambda^{L_{1}}_{\mbox{hinge}}$ (mod 1) & $\lambda^{L_{2}}_{\mbox{hinge}}$ (mod 1/3) & $Q_{\mbox{corner}}$ (mod $2/N$) \\
		\hline

195 & $T$ & $\frac{\Delta b + \Delta d}{2}$ & $\frac{\Delta a + \Delta d}{4}$ & $\frac{\Delta a + \Delta d}{2}$ & \rdelim\}{27}{3mm}\multirow{27}{*}{$\frac{\Delta a}{N}$} \\

196 & $T$ & 0 & $\frac{\Delta a + \Delta b}{4}$ & $\frac{\Delta a + \Delta b}{2}$ &  \\

197 & $T$ & $\frac{\Delta a + \Delta b}{2}$ & $\frac{\Delta a + \Delta b}{4}$ & $\frac{\Delta a + \Delta b}{2}$ &  \\

200 & $T_{h}$ & $\frac{\Delta b + \Delta d}{2}$ & $\frac{\Delta a + \Delta d}{4}$ & $\frac{\Delta a + \Delta d}{2}$ &  \\

201 & $T$ & $\frac{\Delta a + \Delta d}{2}$ & $\frac{\Delta a + \Delta d}{4}$ & 0 &  \\

202 & $T_{h}$ & 0 & $\frac{\Delta a + \Delta b}{4}\equiv \frac{\Delta c + \Delta d}{2}$ & 0 &  \\

203 & $T$ & 0 & $\frac{\Delta a + \Delta b}{4}$ & $\frac{\Delta a + \Delta b}{2}$ &  \\

204 & $T_{h}$ & $\frac{\Delta a + \Delta b}{2}$ & $\frac{\Delta a + \Delta b}{4}$ & 0 &  \\

207 & $O$ & $\frac{\Delta b + \Delta d}{2}$ & $\frac{\Delta a + \Delta d}{4}$ & $\frac{\Delta a + \Delta d}{2}$ &  \\

208 & $T$ & $\frac{\Delta a + \Delta d}{2}$ & $\frac{\Delta a + \Delta d}{4}$ & 0 &  \\

209 & $O$ & 0 & $\frac{\Delta a + \Delta b}{4}\equiv \frac{\Delta c + \Delta d}{2}$ & 0 &  \\

210 & $T$ & 0 & $\frac{\Delta a + \Delta b}{4}$ & 0 &  \\

211 & $O$ & $\frac{\Delta a + \Delta b}{2}$ & $\frac{\Delta a + \Delta b}{4}$ & 0 &  \\

215 & $T_{d}$ & $\frac{\Delta b + \Delta d}{2}$ & $\frac{\Delta a + \Delta d}{4}$ & $\frac{\Delta a + \Delta d}{2}$ &  \\

216 & $T_{d}$ & 0 & $\frac{\Delta a + \Delta b}{4}$ & $\frac{\Delta a + \Delta b}{2}$ &  \\

217 & $T_{d}$ & $\frac{\Delta a + \Delta b}{2}$ & $\frac{\Delta a + \Delta b}{4}$ & 0 &  \\

218 & $T$ & $\frac{\Delta a + \Delta b}{2}$ & $\frac{\Delta a + \Delta b}{4}$ & 0 &  \\

219 & $T$ & 0 & $\frac{\Delta a + \Delta b}{4}$ & 0 &  \\

221 & $O_{h}$ & $\frac{\Delta b + \Delta d}{2}$ & $\frac{\Delta a + \Delta d}{4}$ & $\frac{\Delta a + \Delta d}{2}$ &  \\

222 & $O$ & $\frac{\Delta a + \Delta b}{2}$ & $\frac{\Delta a + \Delta b}{4}$ & 0 &  \\

223 & $T_{h}$ & $\frac{\Delta a + \Delta b}{2}$ & $\frac{\Delta a + \Delta b}{4}$ & 0 &  \\

224 & $T_{d}$ & $\frac{\Delta a + \Delta d}{2}$ & $\frac{\Delta a + \Delta d}{4}$ & 0 &  \\

225 & $O_{h}$ & 0 & $\frac{\Delta a + \Delta b}{4}\equiv \frac{\Delta c + \Delta d}{2}$ & 0 &  \\

226 & $O$ & 0 & $ \frac{\Delta b + \Delta c}{2}$ & 0 &  \\

227 & $T_{d}$ & 0 & $\frac{\Delta a + \Delta b}{4}$ & $\frac{\Delta a + \Delta b}{2}$ &  \\

228 & $T$ & 0 & $\frac{\Delta a + \Delta d}{2}$ & 0 &  \\

229 & $O_{h}$ & $ \frac{\Delta a + \Delta b}{2}$ & $\frac{\Delta a + \Delta b}{4}$ & 0 &  
\end{tabular}
\end{ruledtabular}
\label{Tab:real-space_formulas}
\end{table*}

Here, we discuss how to extend the theory so far to other crystal shapes with other symmetries. 
Before we present our real-space formulas for various cubic crystals, let us make two comments to clarify our setup.

First, we explain the crystal shapes studied in the present paper. 
We focus on vertex-transitive polyhedra. 
In the vertex-transitive polyhedra, all of the corners are connected by symmetry operations. 
Therefore all the corners have the same quantized fractional corner charges. 
Moreover, it is noted that the vertex-transitive polyhedra have large varieties \cite{ROBERTSON197079,
https://doi.org/10.1112/jlms/s2-2.1.125,
bridges2002:320,
polyhedra,
Grunbaum,Leopold}. 
To simplify our discussion, we focus only on the vertex-transitive polyhedra with genus 0 \cite{ROBERTSON197079,
https://doi.org/10.1112/jlms/s2-2.1.125,
bridges2002:320,
polyhedra}, and we do not consider those with higher genus
\cite{Grunbaum,
Leopold}.  
The vertex-transitive polyhedra with genus 0 are further classified into spherical and cylindrical families \cite{https://doi.org/10.1112/jlms/s2-2.1.125,polyhedra}. 
In this paper, we limit ourselves to the spherical family of vertex-transitive polyhedra.
The polyhedra in this family are listed in Refs.~\cite{https://doi.org/10.1112/jlms/s2-2.1.125,polyhedra}. Among them, we need to consider only the polyhedra which are compatible with one of the crystal point groups. It is known that there are families of vertex-transitive polyhedra corresponding to the point groups $O$, $O_h$, $T$, $T_d$, and $T_h$ \cite{https://doi.org/10.1112/jlms/s2-2.1.125,polyhedra}.

\begin{figure*}
    \centering
    \includegraphics[scale = 0.65, angle = 90]{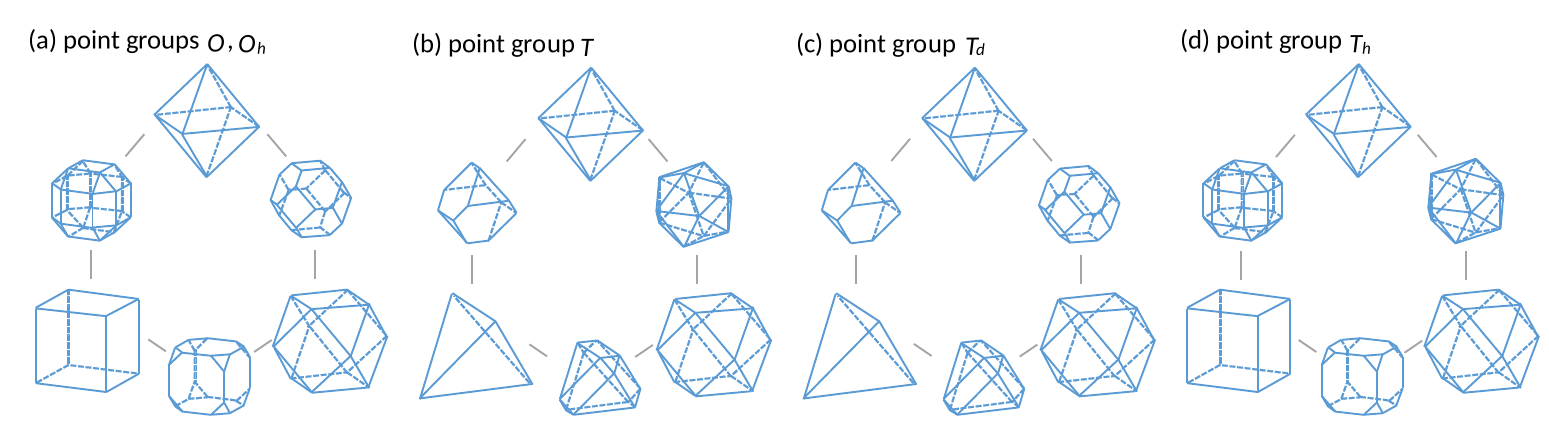}
    \caption{Vertex-transitive polyhedra corresponding to the cubic point group (a) $O$ and $O_{h}$, (b) $T$, (c) $T_{d}$, and (d) $T_{h}$.}
    \label{fig:polyhedra}
\end{figure*}

Among these vertex-transitive polyhedra, some have surfaces whose Miller indices are not fixed but can be continuously deformed. In such cases, 
a continuous change of the surface orientations leads to a continuous change 
in the contribution of the surface polarization to the hinge charge, and obstructs quantization of the hinge charge. Therefore, the hinge charge is not quantized, and we
cannot impose the condition for zero hinge charge, which is necessary to guarantee the corner charge to be well--defined. Thus, in our study, we can exclude such polyhedra, and we deal with the vertex-transitive polyhedra which are spherical and have fixed Miller indices for every surface. 
The 17 types of crystal shapes in 
Fig.~\ref{fig:polyhedra} satisfy such requirements. 

Second, we explain the bulk symmetries. {Since we focus on the crystal shapes 
with point group symmetries $O$, $O_h$, $T$, $T_d$, and $T_h$, }
in this study, we consider the tetrahedral and cubic SGs between the SG numbers 195 and 230. 
Here, SG numbers 198, 199, 205, 206, 212, 213, 214, 220, and 230 are excluded because not all the corners of the corresponding polyhedra for the above SGs are not equivalent and the corner charges are not quantized. 
Therefore, we focus on the 27 tetrahedral and cubic SGs except for the above 9 SGs. 

Through the calculation on the 17 types of crystal shapes, we find that it is enough to calculate for three crystal shapes; a cube, an octahedron and a tetrahedron, because the hinge and corner charges for the other crystal shapes can be calculated from the results obtained from these three crystal shapes through the similar discussion in Sec.~\ref{Sec.2}. 
Thus, we only focus on the cubic, octahedral, and tetrahedral crystal shapes. 
We summarize the real-space formulas obtained by the calculation process similar to that in Sec.~\ref{Sec.2} in Table~\ref{Tab:real-space_formulas}. 
In all the cases, we assume that the center of the crystal is at the Wyckoff position $a$.
Furthermore, since the surface charge densities are coincidentally identical among three crystal shapes (cube, octahedron, and tetrahedron) in terms of modulo 1, we summarize the surface charge density in one column in Table~\ref{Tab:real-space_formulas}. 
We find that, for all the space groups, the hinge charge densities in an octahedron and a tetrahedron are the same in terms of modulo $\abs{e}/3$. 
Thus, we write them together in the column $\lambda^{L_{2}}_{\text{hinge}}$ and the hinge charge density in a cube is written in the column $\lambda^{L_{1}}_{\text{hinge}}$. 
Surprisingly, we also find that the corner charges can be determined only from the information of electrons and ions at the Wyckoff position $1a$ in the bulk unit cell under the charge neutrality conditions for the bulk, surfaces, and hinges for all the space group and the crystal shapes.
Namely, in all the cases, the corner charge is universally given by
\begin{align}
    \label{eq:formula_corner}
    Q_{\text{corner}} = - \frac{\Delta a}{N}\abs{e}\ (\text{mod}\ \frac{2\abs{e}}{N}),
\end{align}
where $N$ represents the number of the corners of the crystal shapes.
A similar result for two-dimensional cases has
been shown in Ref.~\cite{PhysRevB.103.205123}. 
We also note that our corner charge formula is valid only for the crystal shapes where all the corners are connected by any point-group operations at the Wyckoff position $a$ within the space-group operations. 
For example, the cube is not included in Figs.~\ref{fig:polyhedra}(b) and (c), and our formula \eqref{eq:formula_corner} cannot be applied to the cube in SGs where the point group at the Wyckoff position $a$ is either $T$ or $T_{d}$, such as SG No.~195.

\section{Corner charge formulas for various space groups with cubic symmetry in terms of the topological invariant \label{Sec.4}}
In Sec.~\ref{Sec.3}, we find that corner charge formulas are universally equal to the filling anomaly divided by the number of corners. 
In this section, we derive corner charge formulas with all the SGs shown in Table~\ref{Tab:real-space_formulas} in terms of bulk band topology in $k$-space. 
To obtain these formulas for both spinful and spinless systems with time-reversal symmetry (TRS), we use the method of the elementary band representation (EBR) matrix according to Refs.~\cite{PhysRevB.105.045126,PhysRevB.103.165109}. 
In fact, for all the spinful systems and most of the spinless systems in Table~\ref{Tab:real-space_formulas}, we find that corner charge formulas are expressed solely by the number of irreps in the occupied bands at HSPs in $k$-space from the EBR method, and the formulas are shown in Appendix~\ref{App:1}.
In this study, we derive corner charge formulas for the remaining 10 SGs, which cannot be determined only from the EBR method, i.e., SG numbers 195, 196, 197, 201, 203, 207, 208, 209, 210, and 211. 

\subsection{EBR matrix and its ambiguity \label{IVA}}

In this section, we derive a formula to extract the irreps of Wannier orbitals at Wyckoff positions in the real space from a vector $\bm{v}$ of the numbers of irreps at HSPs in the momentum space for a given filled band.
Here, the set of filled bands is assumed to have the trivial symmetry indicator~\cite{Po_2017, SI_SA_Watanabe}. 
Here we explain the EBR matrix \cite{PhysRevB.105.045126,PhysRevB.103.165109}. 
The EBR matrix $M$ is an integer matrix that informs us about the relationship between Wannier orbitals, i.e., irreps at Wyckoff positions in real space and irreps at HSPs in momentum space, as explained below.
Let $\bm{v}$ be a column vector consisting of the number of filled bands with each irrep at each HSPs in momentum space in the filled bands, i.e.,
\begin{align}
    \bm{v} = \left(n_{\Pi}^{\rho_1}, n_{\Pi}^{\rho_2}, \cdots\right)^{t},
\end{align}
where $n_{\Pi}^{\rho}$ represents the number of filled bands with irrep $\rho$ at HSP $\Pi$.
Similarly, let $\bm{n}$ denote a column vector composed of the number of filled Wannier orbitals at each maximal Wyckoff position
\begin{align}
    \bm{n} = \left(a_{\omega}^{\rho_1}, a_{\omega}^{\rho_2}, \cdots\right)^{t},
\end{align}
where $a_{\omega}^{\rho}$ is the number of the Wannier orbital $\rho$ at Wyckoff position $\omega$. 
For an arbitrary AI, we can always write 
\begin{align}
    \bm{v} = M \bm{n}.
    \label{eq:EBR}
\end{align}

To derive momentum-space formulas of the filling anomaly, what we have to do is obtain the number of filled Wannier orbitals at each maximal Wyckoff position for a given $\bm{v}$. This is achieved by the Smith decomposition of $M$:
\begin{align}
    M = U^{-1} \Lambda V^{-1},
    \label{eq:smith}
\end{align}
where $\Lambda$ is an integer diagonal matrix, whose matrix rank is $N$, in the form of $\Lambda_{i j} = \lambda_{i}\delta_{i j}$ ($\lambda_i>0$ for $i=1,\dots,N$ and $\lambda_i=0$ for $i>N$, and $\lambda_i$ divides $\lambda_j$ if $i<j\leq N$), and $U$ and $V$ are unimodular.
The most general solution of Eq.~(\ref{eq:EBR}) is 
\begin{align}
    \bm{n} = V \Lambda^{p} U \bm{v} + V \bm{n}_{0},
    \label{eq:n_vector}
\end{align}
where $\Lambda^{p}$ is the pseudoinverse matrix of $\Lambda$ (i.e., $\Lambda^p$ is the diagonal matrix whose entry is $\lambda_i^{-1}$ for $i=1,\dots, N$ and $0$ for $i>N$), and $\bm{n}_{0}$ is any vector satisfying $\Lambda \bm{n}_{0} = 0$. 
Namely, $n_{0,j\leq N} = 0$, and $n_{0,j>N}$ are arbitrary integers.
The triviality of the symmetry indicator ensures that $\bm{n}$ is an integer vector.
The second term of Eq.~(\ref{eq:n_vector}) represents an ambiguity for the solution of $\bm{n}$ under the given value of $\bm{v}$, and thus, $n_{i}$ can be determined only modulo $\text{gcd}(V_{i j} |_{j > N})$, where gcd indicates the greatest common divisor. 
Therefore, the number of Wannier orbitals at the Wyckoff position $\omega$ is 
\begin{equation}
    \begin{aligned}[b]
        n_{\omega} =& \sum_{i \in \omega} \text{dim}(\rho_{i})(V \Lambda^{p} U \bm{v})_{i} \\
        &\left(\text{mod gcd}\left(\sum_{i \in \omega} \text{dim}(\rho_{i}) V_{i j}\right)_{j > N} \right),
        \label{eq:n_omega}
    \end{aligned}
\end{equation}
where $\rho_{i}$ represents the $i$th irrep of the vector $\bm{n}$, and $\sum_{i \in \omega}$ runs over all the irreps $i$ at the Wyckoff position $\omega$.  
As we learned in Sec.~\ref{Sec.2}, the filling anomaly is given by $\Delta a \equiv n_{a} - m_{a}$ in terms of modulo $2$ to determine the corner charges. 
Therefore, we need to determine the value of $n_{a}$ in Eq.~(\ref{eq:n_omega}) in terms of modulo $2$. 

We note that the formula (\ref{eq:n_omega}) is also applicable to fragile insulators, which have the trivial symmetry indicator. Namely, the vector  $\bm{v}$ for a fragile insulator is given by a difference between those for two atomic insulators, and we can expect that its corner charge should be given by the difference of the corner charges of these atomic insulators, provided that their bulk, surfaces, and hinges are charge neutral. It means that the linear relationship 
(\ref{eq:n_omega}) between $n_\omega$ and $\bm{v}$ holds also for fragile insulators. 

\begin{figure}[H]
    \centering
    \includegraphics[scale = 0.65]{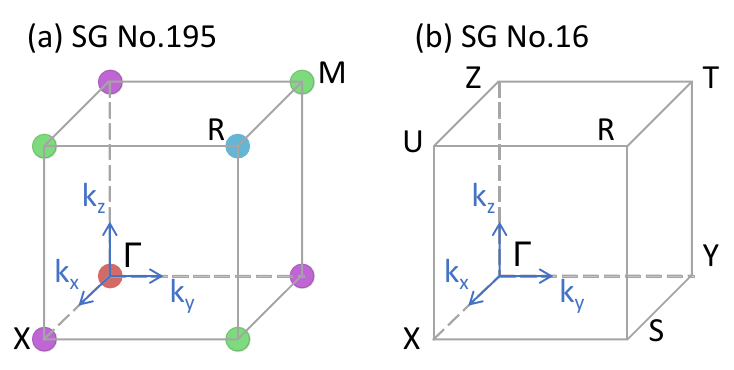}
    \caption{HSPs in the momentum space with (a) SG No.195 and (b) SG No.16. The points with the same color belong to the same k-vector star. The gray cube represents $1/8$ of the Brillouin zone. The $\Gamma$ point is the origin in $k$-space.}
    \label{fig:BZ}
\end{figure}

\subsection{Setup}
We carry out the calculations of Eq.~(\ref{eq:n_omega}) for both of the spinful and spinless cases with TRS in Table~\ref{Tab:real-space_formulas}, and we find that the value of $n_{a}$ from Eq.~(\ref{eq:n_omega}) cannot be determined in terms of modulo 2 for spinless systems with the 10 SGs, of the SG numbers 195, 196, 197, 201, 203, 207, 208, 209, 210, and 211. 
Namely, in these cases, $\text{gcd} \left(\sum_{i \in a} \text{dim}(\rho_{i}) V_{i j}\right)_{j > N}$ is an odd number.

Here we focus on the spinless system with the SG No.~195 as an example, and we will see that the remaining 9 cases can be studied similarly. 
The SG No.~195 $(P23)$ is the simplest space group among cubic and tetrahedral SGs, and is a subgroup of the other cubic and tetrahedral SGs. 
Therefore, if we calculate the filling anomaly for the SG No.~195, we can directly extend it to the other cubic and tetrahedral SGs. 

We show the Wyckoff positions and HSPs in the SG No.~195 in Figs.~\ref{fig:unit_cell} and \ref{fig:BZ} (a), respectively. 
The Wyckoff positions in the unit cell with SG No.~195 are the same as those with SG No.~207. 
The site-symmetry groups of $1a$ and $1b$ are isomorphic to the point group $T$, and those of $3c$ and $3d$ are isomorphic to the point group $D_{2}$. 
On the other hand, there are four kinds of HSPs called $\Gamma, X, M,$ and $R$ in the SG No.~195. 
In Fig.~\ref{fig:BZ} (a), the red, green, purple, and blue points represent the $\Gamma, X, M$ and $R$ points, respectively. 
The little groups at $\Gamma$ and $R$ are isomorphic to the point group $T$ and those at $X$ and $M$ are isomorphic to the point group $D_{2}$. 
The irreps of $T$ and $D_{2}$ are listed in Tables~\ref{tab:T} and  \ref{tab:D2}, respectively. 

\begin{table}[H]
    \caption{Character table of the point group $T$.}
    \centering
    \begin{ruledtabular}
    \begin{tabular}{cccccc}
        $T$ & $E$ & $C_{3(111)}$ & $C_{2x}$ & $C_{2y}$ & $C_{2z}$  \\ \hline
        $A$ & 1 & 1 & 1 & 1 & 1 \\
        $E^{1}$ & 1 & $e^{-i\frac{2\pi}{3}}$ & 1 & 1 & 1 \\
        $E^{2}$ & 1 & $e^{i\frac{2\pi}{3}}$ & 1 & 1 & 1 \\
        $T$ & 3 & 0 & $-1$ & $-1$ & $-1$ 
    \end{tabular}
    \end{ruledtabular}
    \label{tab:T}
\end{table}
\begin{table}[H]
    \caption{Character table of the point group $D_{2}$.}
    \centering
    \begin{ruledtabular}
    \begin{tabular}{ccccc}
         $D_{2}$ & $E$ & $C_{2x}$ & $C_{2y}$ & $C_{2z}$ \\ \hline
         $A_{1}$ & 1 & 1 & 1 & 1 \\
         $B_{1}$ & 1 & $-1$ & $-1$ & 1 \\
         $B_{2}$ & 1 & $-1$ & 1 & $-1$ \\
         $B_{3}$ & 1 & 1 & $-1$ & $-1$ 
    \end{tabular}
    \end{ruledtabular}
    \label{tab:D2}
\end{table}

Next, we consider the EBR matrix for SG No.~195. 
Band representations for the Wannier orbitals located at all the maximal Wyckoff positions are expressed as a vector $\bm{v}$ in the form
\begin{equation}
    \begin{aligned}[b]
        ( n^{E^{2}}_{\Gamma}, n^{E^{1}}_{\Gamma}, n^{A}_{\Gamma},n^{T}_{\Gamma},n^{B_{1}}_{X}, n^{B_{2}}_{X}, n^{B_{3}}_{X}, n^{A_{1}}_{X}, n^{B_{1}}_{M}, \\
        n^{B_{2}}_{M}, n^{B_{3}}_{M}, n^{A_{1}}_{M}, n^{E^{2}}_{R}, n^{E^{1}}_{R}, n^{A}_{R}, n^{T}_{R})^{t},
    \end{aligned}
\end{equation}
where $n^{\rho}_{\Pi}$ indicates the number of times the irrep $\rho$ appears in a given set of filled bands at HSP $\Pi$.
Any vector $\bm{v}$ of the filled band whose symmetry indicator is trivial can be written as a linear combination of the column vectors of the EBR matrix with integer coefficients.
The coefficients form a vector $\bm{n}$ in the form
\begin{equation}
    \begin{aligned}[b]
        (a^{E}_{1a},a^{A}_{1a},a^{T}_{1a},a^{E}_{1b}, a^{A}_{1b},a^{T}_{1b}, a^{B_{1}}_{3c}, a^{B_{2}}_{3c}, \\
        a^{B_{3}}_{3c}, a^{A_{1}}_{3c}, a^{B_{1}}_{3d},a^{B_{2}}_{3d}, a^{B_{3}}_{3d}, a^{A_{1}}_{3d})^{t},
        \label{eq:N}
    \end{aligned}
\end{equation}
where $a^{\rho}_{\omega}$ is the number of atomic insulators whose Wyckoff position is $\omega$ and the irrep is $\rho$ of the site-symmetry group. 
Here, $E = E^1 \oplus E^2$ because of TRS.
In this basis, we construct the EBR matrix for the SG No.~195 by using Topological Quantum Chemistry in the Bilbao Crystallographic Server \cite{Aroyo:xo5013} as follows: 
\begin{widetext}
\begin{align}
    M = \mqty(1&0&0&1&0&0&0&0&0&1&0&0&0&1 \\
              1&0&0&1&0&0&0&0&0&1&0&0&0&1 \\
              0&1&0&0&1&0&0&0&0&1&0&0&0&1 \\
              0&0&1&0&0&1&1&1&1&0&1&1&1&0 \\
              0&0&1&0&0&1&1&0&2&0&0&1&2&0 \\
              0&0&1&2&1&0&1&0&0&2&1&0&1&1 \\
              0&0&1&0&0&1&1&2&0&0&2&1&0&0 \\
              2&1&0&0&0&1&0&1&1&1&0&1&0&2 \\
              0&0&1&2&1&0&2&0&0&1&1&1&1&0 \\
              0&0&1&0&0&1&0&0&2&1&0&0&2&1 \\
              0&0&1&0&0&1&0&2&0&1&2&0&0&1 \\
              2&1&0&0&0&1&1&1&1&0&0&2&0&1 \\
              1&0&0&1&0&0&1&0&0&0&0&1&0&0 \\
              1&0&0&1&0&0&1&0&0&0&0&1&0&0 \\
              0&1&0&0&1&0&1&0&0&0&0&1&0&0 \\
              0&0&1&0&0&1&0&1&1&1&1&0&1&1).
\end{align}
Here, the $(i, j)$ component of $M$ indicates the number of times the $i$ th irrep in the momentum space appears in the EBR induced from the $j$ th irrep in the real space. 
We can obtain the matrices $V$ and $\Lambda$ in Eq.~(\ref{eq:smith}) by carrying out Smith decomposition to this EBR matrix $M$ as follows:
\begin{align}
    V = \mqty(1&0&0&-1&0&0&0&-1&-1&0&0&-1&-1&-1 \\
              0&1&0&-1&0&0&-2&-1&0&-1&-1&-1&-1&-2 \\
              0&0&1&-1&0&0&-1&-2&-2&-1&-1&-1&-3&-2 \\
              0&0&0&0&0&0&-1&0&1&-1&-1&0&0&-1 \\
              0&0&0&0&0&0&1&0&0&0&0&0&0&0 \\
              0&0&0&0&0&0&0&1&0&0&0&0&0&0 \\
              0&0&0&1&1&0&1&1&0&1&1&0&1&2 \\
              0&0&0&0&0&1&0&0&1&0&-1&0&1&0 \\
              0&0&0&0&0&0&0&0&1&0&0&0&0&0 \\
              0&0&0&0&0&0&0&0&0&1&0&0&0&0 \\
              0&0&0&0&0&0&0&0&0&0&1&0&0&0 \\
              0&0&0&0&0&0&0&0&0&0&0&1&0&0 \\
              0&0&0&0&0&0&0&0&0&0&0&0&1&0 \\
              0&0&0&1&0&0&1&1&0&0&1&1&1&2),
              \label{eq:V}
\end{align}
\begin{align}
    \lambda_{1} = \lambda_{2} = \cdots = \lambda_{6} = 1, N = 6.
\end{align}
To determine the corner charge, we need to calculate $\Delta a \equiv n_{1a} - m_{1a}\ (\text{mod}\ 2)$ as explained in Sec.~\ref{Sec.3}. 
Meanwhile, from Eq.~(\ref{eq:V}), the ambiguity of $n_{1a}$ in Eq.~(\ref{eq:n_omega}) is given in terms of modulo 1:
\begin{align}
\text{gcd}\left(\sum_{i =1,2,3} \text{dim}(\rho_{i}) V_{i j}\right)_{j > 6}
&= \gcd \left[(3,1,2)
\mqty(
    0&-1&-1&0&0&-1&-1&-1 \\
    -2&-1&0&-1&-1&-1&-1&-2 \\
    -1&-2&-2&-1&-1&-1&-3&-2 \\
    )\right] \nonumber\\
&= \gcd (-4,-8,-7,-3,-2,-6,-10,-9)=1.
\end{align}
Thus we cannot determine the corner charge formula for SG No.~195 from the EBR matrix alone. 
\end{widetext}

\newpage
\subsection{Corner charge formulas using a new topological invariant}
\label{sec:Corner charge formulas using a new topological invariant}
In order to construct the corner charge formula for the SG No.~195 in terms of the bulk band structure, we need additional information for the occupied eigenstates. 
For this purpose, we utilize the $\Z_2$ invariant, which we call $\xi \in \{0,1\}$, introduced in Ref.~\cite{Ono=Shiozaki_top_inv}.

The $\Z_2$ invariant is defined for the SG No. $P222$ (No.~16) with TRS in spinless electrons, which is a subgroup of the SG No.~195.
The difference between the SGs No.~16 and No.~195 lies in that the SG No.~16 does not have 3-fold rotations. 
There are eight HSPs in the SG No.~16 shown in Fig.~\ref{fig:BZ} (b). 
The little group on high-symmetry lines along the $C_{2\mu \in \{x,y,z\}}$-axis is $\Z_2 = \{E,C_{2\mu}\}$, and that at the HSPs is $D_2$ whose character table is given by Table~\ref{tab:D2}.
We call the trivial and sign representation of $\Z_2$ as $A$ and $B$, respectively.
The symmetry constraints on the Hamiltonian $H(\bk)$ in Bloch-momentum space is summarized as 
\begin{align}
    &V_T(\bk) H(\bk)^* V_T(\bk)^\dag = H(-\bk), \\
    &V_{C_{2\mu}}(\bk) H(\bk) V_{C_{2\mu}}(\bk)^\dag = H(C_{2\mu}\bk), \quad \mu=x,y,z,  \\
    &V_T(-\bk)V_T(\bk)^*=1, \\
    &V_{C_{2\mu}}(C_{2\mu}\bk)V_{C_{2\mu}}(\bk)=1, \quad \mu=x,y,z,  \\
    &V_T(C_{2\mu}\bk)V_{C_{2\mu}}(\bk)^* = V_{C_{2\mu}}(-\bk)V_T(\bk). 
\end{align}
On each high-symmetry line segment $\bm{\Pi}\bm{\Pi}'$ connecting two HSPs $\bm{\Pi}$ and $\bm{\Pi}'$, we introduce a Wilson line $e^{i\gamma^B_{\bm{\Pi}\bm{\Pi}'}}$ of the $B$-irrep as follows.
Let $\Phi^B(\bk) = (\ket{u_1^B(\bk)}, \dots,\ket{u_n^B(\bk)})$ be an orthogonal set of occupation Bloch states with the $B$-irrep on the line segment $\bm{\Pi}\bm{\Pi}'$. 
Here, $n$ is the number of the $B$-irreps, and when $n=0$ we set $e^{i\gamma^B_{\bm{\Pi}\bm{\Pi}'}}=1$. 
The Wilson line $e^{i\gamma^B_{\bm{\Pi}\bm{\Pi}'}}$ is defined as 
\begin{align}
    \gamma^B_{\bm{\Pi}\bm{\Pi}'} = {\rm Arg} \prod_{\bk=\bm{\Pi}}^{\bm{\Pi}'-\delta \bk} \det \left[ \Phi^B(\bk+\delta \bk)^\dag \Phi^B(\bk) \right], 
\end{align}
where $\delta \bk = \epsilon (\bm{\Pi}'-\bm{\Pi}), \epsilon>0$ is a small displacement vector. 
The Wilson line $e^{i\gamma^B_{\bm{\Pi}\bm{\Pi}'}}$ is not gauge invariant as it changes under the gauge transformation at the endpoints $\Phi^B(\bk) \mapsto \Phi^B(\bk) W^B(\bk)$ with $W^B(\bk) \in U(n)$ a unitary matrix for $\bk = \bm{\Pi}, \bm{\Pi}'$. 
We partially fix a gauge of $\Phi^B(\bm{\Pi})$ of all the HSPs $\bm{\Pi}$ as follows: 
At each HSP $\bm{\Pi}$, the occupied Bloch frame is decomposed as $\Phi^{\rho}(\bm{\Pi})$ of four irreps $\rho \in \{A, B_1, B_2, B_3\}$ listed in Table~\ref{tab:D2}. 
In computing the Wilson line of the line segment parallel to the $C_{2x}$, $C_{2y}$, or $C_{2z}$-axis, the Bloch frame $\Phi^B(\bm{\Pi})$ at the endpoint $\bm{\Pi}$ is set as:
\begin{align}
\Phi^B(\bm{\Pi}) = 
\begin{cases} 
(\Phi^{B_1}(\bm{\Pi}), \Phi^{B_2}(\bm{\Pi})) & \text{for } C_{2x}\text{-axis}, \\
(\Phi^{B_1}(\bm{\Pi}), \Phi^{B_3}(\bm{\Pi})) & \text{for } C_{2y}\text{-axis}, \\
(\Phi^{B_2}(\bm{\Pi}), \Phi^{B_3}(\bm{\Pi})) & \text{for } C_{2z}\text{-axis}, 
\end{cases}
\label{eq:gauge_fixed_cond_HSP}
\end{align}
Note that the gauge constraints (\ref{eq:gauge_fixed_cond_HSP}) are numerically implemented straightforwardly. 
The $\Z_2$ invariant is given as~\cite{Ono=Shiozaki_top_inv} 
\begin{widetext}
\begin{equation}
    \begin{aligned}[b]
        (-1)^{\xi} =& e^{i \gamma^B_{\bm{\Gamma}\mathbf{X}}}e^{i \gamma^B_{\mathbf{X}\mathbf{S}}}e^{i \gamma^B_{\mathbf{S}\mathbf{Y}}}e^{i \gamma^B_{\mathbf{Y}\bm{\Gamma}}} \times e^{i \gamma^B_{\mathbf{Z}\mathbf{T}}}e^{i \gamma^B_{\mathbf{T}\mathbf{R}}}e^{i \gamma^B_{\mathbf{R}\mathbf{U}}}e^{i \gamma^B_{\mathbf{U}\mathbf{Z}}} \times e^{i \gamma^B_{\mathbf{Z}\bm{\Gamma}}}e^{i \gamma^B_{\mathbf{X}\mathbf{U}}}e^{i \gamma^B_{\mathbf{Y}\mathbf{T}}}e^{i \gamma^B_{\mathbf{R}\mathbf{S}}} \\
        &\times \prod_{\bm{\Pi} \in \{\bm{\Gamma},\mathbf{S},\mathbf{U},\mathbf{T}\}} \text{det}[(\Phi^{B_{3}}_{\bm{\Pi}})^{\dag}V_{T}(\bm{\Pi})(\Phi^{B_{3}}_{\bm{\Pi}})^{*}]^{-1} \times \prod_{\bm{\Pi} \in \{\mathbf{X},\mathbf{Y},\mathbf{Z},\mathbf{R}\}}\text{det}[(\Phi^{B_{3}}_{\bm{\Pi}})^{\dag}V_{T}(\bm{\Pi})(\Phi^{B_{3}}_{\bm{\Pi}})^{*}] \\
        & \times \exp\left[ -\frac{1}{2}\oint_{\bm{\Gamma}\rightarrow\mathbf{X}\rightarrow\mathbf{S}\rightarrow\mathbf{Y}\rightarrow\bm{\Gamma}} \text{Tr}[(\Phi^{B}(\bm{k}))^{\dag}V_{C_{2z}T}(\bm{k})d V_{C_{2z}T}(\bm{k})^{\dag} \Phi^{B}(\bm{k})] \right] \\
        & \times \exp\left[ \frac{1}{2}\oint_{\mathbf{Z}\rightarrow\mathbf{U}\rightarrow\mathbf{R}\rightarrow\mathbf{T}\rightarrow\mathbf{Z}} \text{Tr}[(\Phi^{B}(\bm{k}))^{\dag}V_{C_{2z}T}(\bm{k})d V_{C_{2z}T}(\bm{k})^{\dag} \Phi^{B}(\bm{k})] \right] \\
        & \times \exp\left[ -\frac{1}{2}(\int_{\mathbf{Z}\rightarrow\bm{\Gamma}} + \int_{\mathbf{X}\rightarrow\mathbf{U}} + \int_{\mathbf{Y}\rightarrow\mathbf{T}} + \int_{\mathbf{R}\rightarrow\mathbf{S}})\text{Tr}[(\Phi^{B}(\bm{k}))^{\dag}V_{C_{2x}T}(\bm{k})d V_{C_{2x}T}(\bm{k})^{\dag} \Phi^{B}(\bm{k})] \right] \times (-1)^{\rho},
        \label{eq:new_invariant}
    \end{aligned}
\end{equation}
Here, $V_{C_{2\mu}T}(\bk) = V_{C_{2\mu}}(-\bk)V_T(\bk)$, and $\rho \in \{0,1\}$ is a quadratic function determined by irreps at HSPs, 
\begin{align}
\rho 
&:= n_{R}^{B_3} \left(n_{S}^{B_3}+n_{T}^{B_2}+n_{T}^{B_3}+n_{U}^{B_2}+n_{U}^{B_3}+n_{Z}^{B_3}\right)
+n_{R}^{B_2} \left(n_{T}^{B_3}+n_{U}^{B_3}+n_{X}^{B_3}+n_{Y}^{B_3}\right) \nonumber \\
&+n_{T}^{B_2} \left(n_{S}^{B_3}+n_{U}^{B_3}+n_{X}^{B_3}\right)+n_{Y}^{B_3} \left(n_{S}^{B_3}+n_{U}^{B_2}+n_{X}^{B_3}+n_{Z}^{B_3}\right)
+n_{S}^{B_3} n_{U}^{B_2}+n_{X}^{B_3} \left(n_{S}^{B_3}+n_{Z}^{B_3}\right)+n_{S}^{B_3} n_{Z}^{B_3}\nonumber \\
&+n_{T}^{B_3} \left(n_{U}^{B_2}+n_{U}^{B_3}+n_{Y}^{B_3}+n_{Z}^{B_3}\right)+n_{U}^{B_3} \left(n_{X}^{B_3}+n_{Z}^{B_3}\right) + n_{S}^{B_3}+n_{T}^{B_3}+n_{U}^{B_3} \quad \mod 2.
\end{align}
\end{widetext}
It is found that $(-1)^\xi$ is gauge invariant under the gauge-fixing condition given by Eq.~(\ref{eq:gauge_fixed_cond_HSP}), is quantized, and satisfies the linearity $\xi(E \oplus F) \equiv \xi(E) + \xi(F)$.

Because SG No.~16 is a subgroup of SG No.~195, we can calculate this $\mathbb{Z}_{2}$ invariant $\xi$ for the AIs in the SG No.~195, corresponding to each irrep at each Wyckoff position. 
We find that the invariant $\xi$ is nontrivial only for the four irreps at the Wyckoff position $1b$: 
\begin{align}
    \xi (a^{A_{1}}_{1b}) &= \xi (a^{B_{1}}_{1b}) = \xi (a^{B_{2}}_{1b}) = \xi (a^{B_{3}}_{1b}) = 1,
    \label{eq:non_trivial_xi}
\end{align}
\begin{align}
    \xi(a^{\rho}_{\omega}) &= 0 \ \text{otherwise}.
\end{align}
 
Then we find that $n_{1a}$ can be determined modulo 2 by incorporating the invariant $\xi$ into the EBR matrix $M$. 
We incorporate $\xi$ into the definition of $\bm{v}$ and define $\Tilde{\bm{v}}$:
\begin{equation}
    \begin{aligned}[b]
        \Tilde{\bm{v}} = ( n^{E^{2}}_{\Gamma}, n^{E^{1}}_{\Gamma}, n^{A}_{\Gamma}, n^{T}_{\Gamma}, n^{B_{1}}_{X}, n^{B_{2}}_{X}, n^{B_{3}}_{X}, n^{A_{1}}_{X}, n^{B_{1}}_{M}, \\
    n^{B_{2}}_{M}, n^{B_{3}}_{M}, n^{A_{1}}_{M}, n^{E^{2}}_{R}, n^{E^{1}}_{R}, n^{A}_{R}, n^{T}_{R}, \xi)^{t}.
    \end{aligned}
\end{equation}
Correspondingly, we define a pseudo-EBR matrix $\Tilde{M}$ from the EBR matrix $M$, by adding one row representing the values of $\xi$ modulo 2. 
Then we apply the Smith decomposition to this pseudo-EBR matrix $\Tilde{M}$, and calculate $n_{1a}$ just as before (see Appendix~\ref{App:2}). 
As a result, we can obtain a formula for $n_{1a}$ (mod 2). 
Then $\Delta a \equiv n_{1a} - m_{1a}\ (\text{mod}\ 2)$ and the corner charge formula for SG No.~195 in terms of the bulk band structure and bulk ionic positions are 
\begin{align}
    \Delta a = n^{A}_{\Gamma} + n^{B_{2}}_{M} + \xi - m_{a}\ (\text{mod}\ 2),
\end{align}
\begin{align}
    Q_{\text{corner}} = - \frac{n^{A}_{\Gamma} + n^{B_{2}}_{M} + \xi - m_{1a}}{N} \abs{e} \ (\text{mod}\ \frac{2\abs{e}}{N}).
\end{align}
We also calculate the corner charge formulas for the remaining nine SGs using $\xi$ in the same way, and these results of the filling anomaly are shown in Table~\ref{tab:corner_charge}.  
The corner charge formulas are given by dividing their filling anomalies by the number of corners, $N$.
\begin{table}[H]
    \caption{The filling anomaly for the 10 SGs, for which the corner charges cannot be determined by the EBR matrix alone. The corner charge formulas are given by dividing their filling anomaly by $N$.}
    \centering
    \renewcommand{\arraystretch}{1.2}
    \begin{ruledtabular}
    \begin{tabular}{cc}
        SG number & filling anomaly $\Delta a$ (mod\ 2)  \\
        \hline
        195 & $n^{A}_{\Gamma} + n^{B_{2}}_{M} + \xi - m_{a}$ \\
        196 & $n^{1}_{X} + n^{4}_{\Gamma} + \xi - m_{a}$ \\
        197 & $\xi - m_{a}$ \\
        201 & $n^{A}_{\Gamma} + n^{B_{2}}_{M} + \xi - m_{a}$ \\
        203 & $n^{1+}_{\Gamma} + n^{4+}_{\Gamma} + \xi - m_{a}$ \\
        207 & $n^{4}_{\Gamma} + n^{5}_{\Gamma} + n^{1}_{R} + n^{2}_{R} + \xi - m_{a}$ \\
        208 & $\xi - m_{a}$ \\
        209 & $n^{1}_{\Gamma} + n^{2}_{\Gamma} + n^{4}_{\Gamma} + n^{5}_{\Gamma} + \xi - m_{a}$ \\
        210 &$n^{2}_{\Gamma} + n^{3}_{\Gamma} + n^{4}_{\Gamma} + \xi - m_{a}$  \\
        211 & $\xi - m_{a}$ 
    \end{tabular}
    \end{ruledtabular}
    \label{tab:corner_charge}
\end{table}

\subsection{Corner charge and TRS}
Although the $\Z_2$ invariant $\xi$ successfully captures the corner charges, it is somewhat puzzling that its definition requires TRS, which seems unrelated to the corner charge. 
In this section, we demonstrate that in the absence of TRS, the $\xi=1$ state can be adiabatically connected to the $\xi=0$ state while keeping a bulk energy gap.

\begin{figure}[H]
    \centering
    \includegraphics[width=\linewidth, trim=0cm 0cm 0cm 0cm]{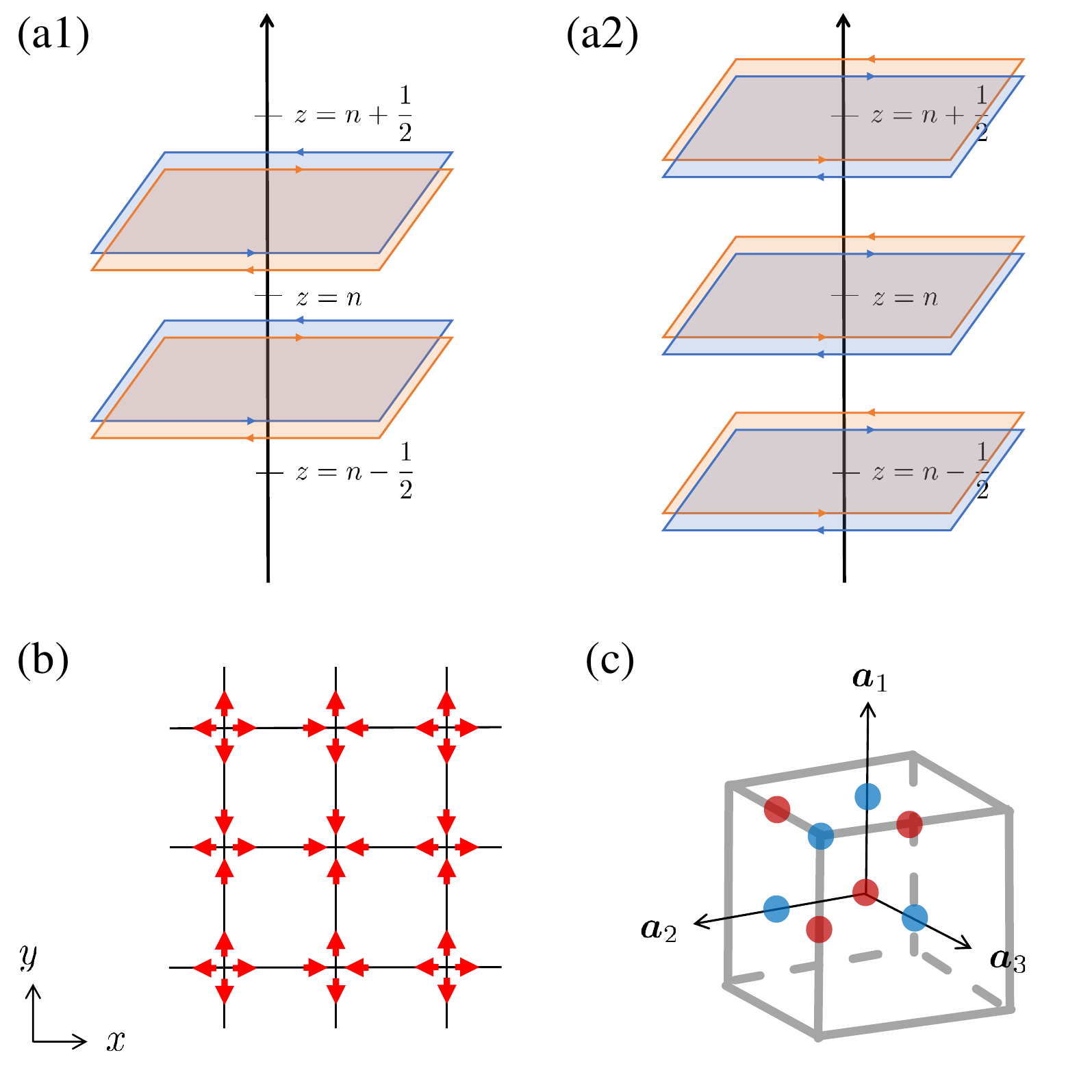}
    \caption{
    (a1) Adiabatic pair creation of Chern insulators on planes $\Z \in Z \pm z_0$. 
    The blue and red layers show Chern insulators with Chern numbers $1$ and $-1$, respectively.
    (a2) Parallel translation of Chern insulators to the planes at $z \in \Z+0$ and $z\in \Z+1/2$, consistent with $P222$ symmetry.
    (b) A vortex structure that is compatible with $P222$ symmetry on the $z=0$ plane. 
    The vector $(M_1, M_2)$ is indicated by the red angles. 
    (c) Illustration of bound states adiabatically created. 
    The figure shows a unit cell, and $\bm{a}_1, \bm{a}_2$, and $\bm{a}_3$ are lattice vectors.
    Red and blue circles represent negative and positive energy-bound states, respectively. 
    }
    \label{fig:d3}
\end{figure}

We start by examining SG No.16 ($P222$) without TRS and incorporate the 3-fold rotation symmetry around the $(111)$-axis later. 
On the layers $z = n + z_0$ with $n \in \Z$ and $z_0 \in (0,1/2)$, we can adiabatically generate two Chern insulators, $H^{(+)}_{z = n +z_0}$ and $H^{(-)}_{z=n+z_0}$ with opposite Chern numbers from vacuum. 
See Fig.~\ref{fig:d3}(a1).
These Hamiltonians are given by
\begin{align}
    H^{(\pm)}_{z = n+z_0}(k_x,k_y) 
    &= k_x \s_x + k_y \s_y \pm m\s_z.
\end{align}
They are invariant under the $C_{2z}$-symmetry,
\begin{align}
    &V_{C_{2z}} H^{(\pm)}_{z = n + z_0}(k_x,k_y) V_{C_{2z}}^\dag 
    = H^{(\pm)}_{z =n +z_0}(-k_x,-k_y)
\end{align}
with $V_{C_{2z}} = \s_z$ and also preserve lattice translation symmetry. 
On the layer $z = n -z_0$, we set the pair of Hamiltonians $\tilde H^{\mp}_{z=n-z_0}(k_x,k_y)$ by $C_{2x}$-rotation as
\begin{align}
\tilde H^{\mp}_{z=n-z_0}(k_x,k_y) &= H^{\pm}_{z =-n+z_0}(k_x,-k_y).    
\end{align}

Next, we adiabatically deform the localized positions of Chern insulators while keeping $P222$ symmetry as:
\begin{align}
    &H^{(+)}_{z=n+z_0} \to H^{(+)}_{z=n+1/2}, \\
    &H^{(-)}_{z=n+z_0} \to H^{(-)}_{z=n+0}, \\
    &\tilde H^{(+)}_{z=n-z_0} \to \tilde H^{(+)}_{z=n-0}, \\
    &\tilde H^{(-)}_{z=n-z_0} \to \tilde H^{(-)}_{z=n-1/2}.
\end{align}
See Fig.~\ref{fig:d3}(a2). 
Consequently, on the $z=n$ layers, we obtain two Chern insulators with opposite Chern numbers:
\begin{align}
    &H^{(-)}_{z=n+0} \oplus \tilde H^{(+)}_{z=n-0}
    = k_x \s_x + k_y \s_y \tau_z - m \s_z, \\
    &V_{C_{2z}}=\s_z, \quad V_{C_{2x}} = \tau_x. 
\end{align}
Here, $\tau_\mu$ denotes the Pauli matrices for the two layers of Chern insulators. 
A generic mass term for \( H^{(-)}_{z=n+0} \oplus \tilde H^{(+)}_{z=n-0} \) is expressed as
\begin{align}
    M_1(x,y)\sigma_y\tau_x + M_2(x,y)\sigma_y\tau_y.
\end{align}
The \( D_2 \) symmetry imposes constraints on the mass terms:
\begin{align}
    &M_1(-x,-y)=-M_1(x,y), \quad M_1(x,-y)=M_1(x,y), \\
    &M_2(-x,-y)=-M_2(x,y), \quad M_2(x,-y)=-M_2(x,y),
\end{align}
which implies that around each of the high-symmetry points \( (x,y) \in (\Z,\Z), (\Z+1/2,\Z), (\Z,\Z+1/2), (\Z+1/2,\Z+1/2) \), the winding number \( W = \frac{1}{2\pi} \oint d\ \text{arg}(M_1+iM_2) \) is an odd integer, \( W \in 2\Z+1 \). 
Then, there appears a low-energy bound state at each high-symmetry point and its energy depends on whether \( W \in 4\Z+1 \) or \( W \in 4\Z-1 \). 
For \( W= \pm 1 \), by taking $M_1\sim x$ and $M_2\sim\pm y$ near $(x,y)=(0,0)$ as an example, we see that the bound state wave functions are given by
\begin{align}
    &\phi^{B_3}_{W=1} \propto \begin{pmatrix} 0 \\ 1 \end{pmatrix}_\sigma \otimes \begin{pmatrix} 1 \\ 1 \end{pmatrix}_\tau
    e^{-(x^2+y^2)/2}, \quad E = m, \\
    &\phi^{B_1}_{W=-1} \propto \begin{pmatrix} 1 \\ 0 \end{pmatrix}_\sigma \otimes \begin{pmatrix} 1 \\ -1 \end{pmatrix}_\tau e^{-(x^2+y^2)/2}, \quad E=-m,
\end{align}
where \( \phi^{B_3}_{W=1} \) and \( \phi^{B_1}_{W=-1} \) represent the \( B_3 \) and \( B_1 \) irreps, respectively. 
Fig.~\ref{fig:d3}(b) illustrates a vortex structure of \( M_1, M_2 \), assuming the absence of vortices except at the high-symmetry points. 
The situation is analogous for the \( z=n+1/2 \) planes. 
Fig.~\ref{fig:d3}(c) provides a possible configuration of positive and negative energy-bound states, where red circles represent occupied Wannier orbitals with charge \( -|e| \), while blue circles are ions. 
The state shown in Fig.~\ref{fig:d3}(c) is TRS invariant and has \( \xi=1 \), which implies that the vacuum state ($\xi=0$) and the $\xi=1$ state are adiabatically connected to each other if TRS is broken. 

The adiabatic deformation described above does not respect the symmetry constraint of SG No.195. 
However, SG No.195 can be enforced by creating three copies using the rotation \(C_3\) along the \((111)\)-axis. 
For the resultant state, we have \((\Delta a, \Delta b, \Delta c, \Delta d) = (3,-3,3,-3)\), which gives the filling anomaly according to Eq.~(\ref{eq:cubic_filling_anomaly}) as \(\eta_n^{\rm perfect,\ type\ 1}=-3\), indicating a nontrivial corner charge. 
We have demonstrated that insulators possessing a corner charge for SG No.195 can be adiabatically connected into insulators lacking a corner charge if TRS is broken. 
Thus, to characterize insulators as a bulk phase based on the corner charge, TRS is crucial.

We note that the adiabatic creation of paired Chern insulators, followed by their transition and subsequent annihilation except at high-symmetry points, can be viewed as a model for the third differential in the real-space Atiyah-Hirzebruch spectral sequence~\cite{10.1093/ptep/ptad086}.

\section{conclusion \label{Sec.5}}
In this paper, we derived formulas for quantized fractional corner charges for three-dimensional insulators with various crystal shapes. 

We derived the formulas in terms of the real-space Wyckoff positions and those in terms of the irreps in $k$-space.
For the quantization of the corner charge, the crystal shape should be a vertex-transitive polyhedron. 
We focused on the crystal shapes of the spherical family of the vertex-transitive polyhedra, corresponding to the five cubic point groups such as $O,O_{h},T,T_{d}$, and $T_{h}$. 
We calculated the real space formulas of corner charges for each crystal shape of these polyhedra, whose center is located at the Wyckoff position $a$. 
Surprisingly, we found that the filling anomaly is given by $\Delta a \equiv n_{1a} - m_{1a}\ (\text{mod}\ 2)$ irrespective of the crystal shapes and the space group considered, where $n_{1a}$ is the number of Wannier orbitals at the Wyckoff position $a$ in the occupied states and $m_{1a}$ is the total charge of ions measured in the unit of $\abs{e}$ at the Wyckoff position $a$.  
As a result, the corner charge formulas in these cases are given by $-\Delta a \abs{e}/N$ where $N$ is the number of corners of the crystal shapes. 

Finally, we took the method of the EBR matrix to obtain the corner charge formulas for the tetrahedral and cubic SGs in terms of the bulk band structures in $k$-space. 
As a result, we found that the corner charge formulas for all the spinful cases and some of the spinless cases are given in terms of the number of occupied irreps at HSPs in $k$-space. 
Meanwhile, in the remaining cases of spinless systems for the 10 SGs are not determined from the EBR matrix alone, because some band structures with different Wyckoff positions share the same irreps at all the HSPs in the BZ. 
Such SG numbers are  195, 196, 197, 201, 203, 207, 208, 209, 210, and 211. 
To solve this problem, we introduced the new $\mathbb{Z}_{2}$ topological invariant $\xi$, which is defined in SG No.~16. 
By incorporating this invariant $\xi$, we constructed the corner charge formulas in terms of bulk band structures in $k$-space for the above the 10 SGs.

\begin{acknowledgments}
H.~W.~and K.~N.~contributed equally to this work. We thank the YITP workshop YITP-T22-02 on ``Novel Quantum States in Condensed Matter 2022", which was helpful in completing this work.
This work was supported by Japan Society for the Promotion of Science (JSPS) KAKENHI Grants No.~JP22K18687, and No.~JP22H00108.
KS was supported by JST CREST Grant No. JPMJCR19T2, and JSPS KAKENHI Grant No. 22H05118 and 23H01097. 
SO thanks RIKEN Special Postdoctoral Researchers Program.
\end{acknowledgments}

\appendix
\section{CORNER CHARGE FORMULAS WITH BOTH SPINFUL AND SPINLESS SYSTEMS FOR THE CUBIC SGs \label{App:1}}
Here we summarize the corner charge formulas for the 27 tetrahedral and cubic SGs for both spinful and spinless systems with TRS. 
We show the filling anomalies $\Delta a_{\text{spinful}}$ for spinful systems and $\Delta a_{\text{spinless}}$ for spinless systems for the above SGs in Table~\ref{tab:corner_spinful_spinless}. 
The corner charge formulas are given by dividing their filling anomalies by $N$. 
We note that all the irreps of each double group considered have even dimensions in the spinful cases due to the Kramers degeneracy \cite{PhysRevResearch.1.033074}. 
Therefore, the corner charges have no electronic contribution for spinful systems. 

\begin{table*}
    \caption{Summary for the filling anomaly for both spinful and spinless systems for the tetrahedral and cubic SGs. The corner charge formulas are given by dividing their filling anomaly by $N$. We note that the filling anomaly for the spinful systems are universally given by $-m_{a}$.}
    \centering
    \renewcommand{\arraystretch}{1.2}
    \begin{ruledtabular}
    \begin{tabular}{ccc}
        SG number & $\Delta a_{\text{spinful}}\ (\text{mod}\ 2)$ & $\Delta a_{\text{spinless}}\ (\text{mod}\ 2)$ \\ \hline
        195 & & $n^{A}_{\Gamma} + n^{B_{2}}_{M} + \xi - m_{a}$ \\ 
        196 & & $n^{1}_{X} + n^{4}_{\Gamma} + \xi - m_{a}$ \\ 
        197 & & $\xi - m_{a}$ \\ 
        200 &  & $n^{4-}_{\Gamma} + n^{4+}_{R} + n^{1-}_{M} + n^{1+}_{X} - m_{a} $ \\ 
        201 & & $\xi - m_{a}$ \\ 
        202 & & $n^{1-}_{\Gamma} + n^{4+}_{\Gamma} + n^{4-}_{\Gamma} + n^{2+}_{X} + n^{1+}_{L} - m_{a}$ \\ 
        203 & & $n^{1+}_{\Gamma} + n^{4+}_{\Gamma} + \xi - m_{a}$ \\ 
        204 & & $n^{4-}_{\Gamma} + n^{1+}_{H} + n^{1}_{P} + n^{1+}_{N} - m_{a}$\\ 
        207 & & $n^{4}_{\Gamma} + n^{5}_{\Gamma} + n^{1}_{R} + n^{2}_{R} + \xi - m_{a}$ \\ 
        208 & & $\xi - m_{a}$ \\ 
        209 & & $n^{1}_{\Gamma} + n^{2}_{\Gamma} + n^{4}_{\Gamma} + n^{5}_{\Gamma} + \xi - m_{a}$ \\ 
        210 & & $n^{2}_{\Gamma} + n^{3}_{\Gamma} + n^{4}_{\Gamma} + \xi - m_{a}$ \\ 
        211 & & $\xi - m_{a}$ \\ 
        215 & $-m_{a}$ & $n^{1}_{\Gamma} + n^{4}_{\Gamma} + n^{5}_{\Gamma} + n^{1}_{R} + n^{2}_{M} + n^{1}_{X} - m_{a}$\\ 
        216 & & $n^{4}_{\Gamma} + n^{2}_{X} + n^{1}_{W} - m_{a}$\\ 
        217 & & $n^{1}_{\Gamma} + n^{4}_{\Gamma} + n^{5}_{\Gamma} + n^{1}_{H} + n^{1}_{P} + n^{2}_{P} - m_{a}$ \\ 
        218 & & $n^{2}_{\Gamma} + n^{1}_{M} - m_{a}$ \\ 
        219 & & $n^{4}_{\Gamma} + n^{1}_{X} - m_{a}$ \\ 
        221 & & $n^{1+}_{\Gamma} + n^{2+}_{\Gamma} + n^{1-}_{R} + n^{2-}_{R} + n^{5-}_{M} + n^{5+}_{X} - m_{a}$\\
        222 & & $n^{1+}_{\Gamma} + n^{1-}_{\Gamma} + n^{4+}_{\Gamma} + n^{5+}_{\Gamma} + n^{1}_{R} - m_{a}$ \\ 
        223 & & $n^{1+}_{\Gamma} + n^{2-}_{\Gamma} + n^{1-}_{M} + n^{2+}_{M} - m_{a}$ \\ 
        224 & & $n^{2-}_{\Gamma} + n^{3-}_{\Gamma} + n^{4-}_{\Gamma} + n^{2+}_{R} + n^{3+}_{R} + n^{4+}_{R} - m_{a}$ \\ 
        225 & & $n^{1+}_{\Gamma} + n^{2-}_{\Gamma} + n^{1+}_{X} + n^{2+}_{X} + n^{3+}_{X} + n^{3-}_{X} + n^{1+}_{L} + n^{1-}_{L} - m_{a}$ \\ 
        226 & & $n^{1+}_{\Gamma} + n^{2+}_{\Gamma} + n^{1+}_{X} + n^{2+}_{X} - m_{a}$ \\ 
        227 & & $n^{1+}_{\Gamma} + n^{1-}_{\Gamma} + n^{2-}_{\Gamma} + n^{3+}_{\Gamma} + n^{4+}_{\Gamma} + n^{1}_{W} - m_{a}$ \\ 
        228 & & $n^{2+}_{\Gamma} + n^{3+}_{\Gamma} + n^{4+}_{\Gamma} - m_{a}$ \\ 
        229 & & $n^{1+}_{\Gamma} + n^{2-}_{\Gamma} + n^{1-}_{H} + n^{2+}_{H} + n^{4}_{P} + n^{5}_{P} - m_{a}$ \\ 
    \end{tabular}
    \end{ruledtabular}
    \label{tab:corner_spinful_spinless}
\end{table*}

\section{SMITH NORMAL FORM OF THE
PSEUDO-EBR MATRIX \texorpdfstring{$\Tilde{M}$}{TEXT} \label{App:2}}
In Sec.~\ref{Sec.4}, we obtained the pseudo-EBR matrix $\Tilde{M}$ from the EBR matrix $M$ by adding a row $(0,0,0,0,1,1$, $0,0,0,0,0,0,0,0)$ in spinless systems with the SG No.~195. 
This new row represents the values of $\xi$ modulo 2 for each atomic insulator represented by Eq.~(\ref{eq:N}) obtained from Eq.~(\ref{eq:non_trivial_xi}). 
As a result, the pseudo-EBR matrix $\Tilde{M}$ is 
\begin{widetext}
\begin{align}
    \Tilde{M} = \mqty(1&0&0&1&0&0&0&0&0&1&0&0&0&1 \\
                      1&0&0&1&0&0&0&0&0&1&0&0&0&1 \\
                      0&1&0&0&1&0&0&0&0&1&0&0&0&1 \\
                      0&0&1&0&0&1&1&1&1&0&1&1&1&0 \\
                      0&0&1&0&0&1&1&0&2&0&0&1&2&0 \\
                      0&0&1&2&1&0&1&0&0&2&1&0&1&1 \\
                      0&0&1&0&0&1&1&2&0&0&2&1&0&0 \\
                      2&1&0&0&0&1&0&1&1&1&0&1&0&2 \\
                      0&0&1&2&1&0&2&0&0&1&1&1&1&0 \\
                      0&0&1&0&0&1&0&0&2&1&0&0&2&1 \\
                      0&0&1&0&0&1&0&2&0&1&2&0&0&1 \\
                      2&1&0&0&0&1&1&1&1&0&0&2&0&1 \\
                      1&0&0&1&0&0&1&0&0&0&0&1&0&0 \\
                      1&0&0&1&0&0&1&0&0&0&0&1&0&0 \\
                      0&1&0&0&1&0&1&0&0&0&0&1&0&0 \\
                      0&0&1&0&0&1&0&1&1&1&1&0&1&1 \\
                      0&0&0&0&1&1&0&0&0&0&0&0&0&0 ).
\end{align}
We can obtain the matrices $\Tilde{V}$ and $\Tilde{\Lambda}$ by using Smith decomposition to this EBR matrix $\Tilde{M}$. 
Accordingly, we can determine $n_{\omega}$ modulo 2 from symmetry indicators and the value of $\xi$: 
\begin{align}
    \Tilde{V} = \mqty(1&0&0&0&0&0&0&-1&0&-1&0&0&0&-1 \\
                      0&1&0&0&-1&0&-1&0&0&-1&1&0&0&-1 \\
                      0&0&1&1&0&-1&0&0&-2&-1&-1&0&-2&-1 \\
                      0&0&0&0&0&0&0&1&0&0&0&0&0&0 \\
                      0&0&0&0&1&0&1&0&0&0&-1&0&0&0 \\
                      0&0&0&0&0&0&0&0&0&0&1&0&0&0 \\
                      0&0&0&0&0&1&0&0&0&1&0&-1&0&1 \\
                      0&0&0&-1&0&0&1&2&-1&2&-2&-1&0&1 \\
                      0&0&0&0&0&0&0&0&1&0&0&0&0&0  \\
                      0&0&0&0&0&0&0&0&0&1&0&0&0&0 \\
                      0&0&0&0&0&0&-1&-2&2&-2&2&1&1&-1 \\
                      0&0&0&0&0&0&0&0&0&0&0&1&0&0 \\
                      0&0&0&0&0&0&0&0&0&0&0&0&1&0 \\
                      0&0&0&0&0&0&0&0&0&0&0&0&0&1 ),
\end{align}
\begin{align}
    \lambda_{1} = \lambda_{2} = \cdots = \lambda_{7} = 1, N = 7.
\end{align}
\end{widetext}

\bibliography{reference}

\begin{thebibliography}{58}%
\makeatletter
\providecommand \@ifxundefined [1]{%
 \@ifx{#1\undefined}
}%
\providecommand \@ifnum [1]{%
 \ifnum #1\expandafter \@firstoftwo
 \else \expandafter \@secondoftwo
 \fi
}%
\providecommand \@ifx [1]{%
 \ifx #1\expandafter \@firstoftwo
 \else \expandafter \@secondoftwo
 \fi
}%
\providecommand \natexlab [1]{#1}%
\providecommand \enquote  [1]{``#1''}%
\providecommand \bibnamefont  [1]{#1}%
\providecommand \bibfnamefont [1]{#1}%
\providecommand \citenamefont [1]{#1}%
\providecommand \href@noop [0]{\@secondoftwo}%
\providecommand \href [0]{\begingroup \@sanitize@url \@href}%
\providecommand \@href[1]{\@@startlink{#1}\@@href}%
\providecommand \@@href[1]{\endgroup#1\@@endlink}%
\providecommand \@sanitize@url [0]{\catcode `\\12\catcode `\$12\catcode
  `\&12\catcode `\#12\catcode `\^12\catcode `\_12\catcode `\%12\relax}%
\providecommand \@@startlink[1]{}%
\providecommand \@@endlink[0]{}%
\providecommand \url  [0]{\begingroup\@sanitize@url \@url }%
\providecommand \@url [1]{\endgroup\@href {#1}{\urlprefix }}%
\providecommand \urlprefix  [0]{URL }%
\providecommand \Eprint [0]{\href }%
\providecommand \doibase [0]{https://doi.org/}%
\providecommand \selectlanguage [0]{\@gobble}%
\providecommand \bibinfo  [0]{\@secondoftwo}%
\providecommand \bibfield  [0]{\@secondoftwo}%
\providecommand \translation [1]{[#1]}%
\providecommand \BibitemOpen [0]{}%
\providecommand \bibitemStop [0]{}%
\providecommand \bibitemNoStop [0]{.\EOS\space}%
\providecommand \EOS [0]{\spacefactor3000\relax}%
\providecommand \BibitemShut  [1]{\csname bibitem#1\endcsname}%
\let\auto@bib@innerbib\@empty
\bibitem [{\citenamefont {Fu}\ \emph {et~al.}(2007)\citenamefont {Fu},
  \citenamefont {Kane},\ and\ \citenamefont {Mele}}]{PhysRevLett.98.106803}%
  \BibitemOpen
  \bibfield  {author} {\bibinfo {author} {\bibfnamefont {L.}~\bibnamefont
  {Fu}}, \bibinfo {author} {\bibfnamefont {C.~L.}\ \bibnamefont {Kane}},\ and\
  \bibinfo {author} {\bibfnamefont {E.~J.}\ \bibnamefont {Mele}},\ }\bibfield
  {title} {\bibinfo {title} {Topological insulators in three dimensions},\
  }\href {https://doi.org/10.1103/PhysRevLett.98.106803} {\bibfield  {journal}
  {\bibinfo  {journal} {Phys. Rev. Lett.}\ }\textbf {\bibinfo {volume} {98}},\
  \bibinfo {pages} {106803} (\bibinfo {year} {2007})}\BibitemShut {NoStop}%
\bibitem [{\citenamefont {Kane}\ and\ \citenamefont
  {Mele}(2005{\natexlab{a}})}]{PhysRevLett.95.146802}%
  \BibitemOpen
  \bibfield  {author} {\bibinfo {author} {\bibfnamefont {C.~L.}\ \bibnamefont
  {Kane}}\ and\ \bibinfo {author} {\bibfnamefont {E.~J.}\ \bibnamefont
  {Mele}},\ }\bibfield  {title} {\bibinfo {title} {${Z}_{2}$ topological order
  and the quantum spin hall effect},\ }\href
  {https://doi.org/10.1103/PhysRevLett.95.146802} {\bibfield  {journal}
  {\bibinfo  {journal} {Phys. Rev. Lett.}\ }\textbf {\bibinfo {volume} {95}},\
  \bibinfo {pages} {146802} (\bibinfo {year} {2005}{\natexlab{a}})}\BibitemShut
  {NoStop}%
\bibitem [{\citenamefont {Kane}\ and\ \citenamefont
  {Mele}(2005{\natexlab{b}})}]{PhysRevLett.95.226801}%
  \BibitemOpen
  \bibfield  {author} {\bibinfo {author} {\bibfnamefont {C.~L.}\ \bibnamefont
  {Kane}}\ and\ \bibinfo {author} {\bibfnamefont {E.~J.}\ \bibnamefont
  {Mele}},\ }\bibfield  {title} {\bibinfo {title} {Quantum spin hall effect in
  graphene},\ }\href {https://doi.org/10.1103/PhysRevLett.95.226801} {\bibfield
   {journal} {\bibinfo  {journal} {Phys. Rev. Lett.}\ }\textbf {\bibinfo
  {volume} {95}},\ \bibinfo {pages} {226801} (\bibinfo {year}
  {2005}{\natexlab{b}})}\BibitemShut {NoStop}%
\bibitem [{\citenamefont {Fu}\ and\ \citenamefont
  {Kane}(2007)}]{PhysRevB.76.045302}%
  \BibitemOpen
  \bibfield  {author} {\bibinfo {author} {\bibfnamefont {L.}~\bibnamefont
  {Fu}}\ and\ \bibinfo {author} {\bibfnamefont {C.~L.}\ \bibnamefont {Kane}},\
  }\bibfield  {title} {\bibinfo {title} {Topological insulators with inversion
  symmetry},\ }\href {https://doi.org/10.1103/PhysRevB.76.045302} {\bibfield
  {journal} {\bibinfo  {journal} {Phys. Rev. B}\ }\textbf {\bibinfo {volume}
  {76}},\ \bibinfo {pages} {045302} (\bibinfo {year} {2007})}\BibitemShut
  {NoStop}%
\bibitem [{\citenamefont {Fu}\ and\ \citenamefont
  {Kane}(2006)}]{PhysRevB.74.195312}%
  \BibitemOpen
  \bibfield  {author} {\bibinfo {author} {\bibfnamefont {L.}~\bibnamefont
  {Fu}}\ and\ \bibinfo {author} {\bibfnamefont {C.~L.}\ \bibnamefont {Kane}},\
  }\bibfield  {title} {\bibinfo {title} {Time reversal polarization and a
  ${Z}_{2}$ adiabatic spin pump},\ }\href
  {https://doi.org/10.1103/PhysRevB.74.195312} {\bibfield  {journal} {\bibinfo
  {journal} {Phys. Rev. B}\ }\textbf {\bibinfo {volume} {74}},\ \bibinfo
  {pages} {195312} (\bibinfo {year} {2006})}\BibitemShut {NoStop}%
\bibitem [{\citenamefont {Thouless}(1983)}]{PhysRevB.27.6083}%
  \BibitemOpen
  \bibfield  {author} {\bibinfo {author} {\bibfnamefont {D.~J.}\ \bibnamefont
  {Thouless}},\ }\bibfield  {title} {\bibinfo {title} {Quantization of particle
  transport},\ }\href {https://doi.org/10.1103/PhysRevB.27.6083} {\bibfield
  {journal} {\bibinfo  {journal} {Phys. Rev. B}\ }\textbf {\bibinfo {volume}
  {27}},\ \bibinfo {pages} {6083} (\bibinfo {year} {1983})}\BibitemShut
  {NoStop}%
\bibitem [{\citenamefont {König}\ \emph {et~al.}(2007)\citenamefont {König},
  \citenamefont {Wiedmann}, \citenamefont {Brüne}, \citenamefont {Roth},
  \citenamefont {Buhmann}, \citenamefont {Molenkamp}, \citenamefont {Qi},\ and\
  \citenamefont {Zhang}}]{doi:10.1126/science.1148047}%
  \BibitemOpen
  \bibfield  {author} {\bibinfo {author} {\bibfnamefont {M.}~\bibnamefont
  {König}}, \bibinfo {author} {\bibfnamefont {S.}~\bibnamefont {Wiedmann}},
  \bibinfo {author} {\bibfnamefont {C.}~\bibnamefont {Brüne}}, \bibinfo
  {author} {\bibfnamefont {A.}~\bibnamefont {Roth}}, \bibinfo {author}
  {\bibfnamefont {H.}~\bibnamefont {Buhmann}}, \bibinfo {author} {\bibfnamefont
  {L.~W.}\ \bibnamefont {Molenkamp}}, \bibinfo {author} {\bibfnamefont {X.-L.}\
  \bibnamefont {Qi}},\ and\ \bibinfo {author} {\bibfnamefont {S.-C.}\
  \bibnamefont {Zhang}},\ }\bibfield  {title} {\bibinfo {title} {Quantum spin
  hall insulator state in hgte quantum wells},\ }\href
  {https://doi.org/10.1126/science.1148047} {\bibfield  {journal} {\bibinfo
  {journal} {Science}\ }\textbf {\bibinfo {volume} {318}},\ \bibinfo {pages}
  {766} (\bibinfo {year} {2007})}\BibitemShut {NoStop}%
\bibitem [{\citenamefont {Bernevig}\ \emph {et~al.}(2006)\citenamefont
  {Bernevig}, \citenamefont {Hughes},\ and\ \citenamefont
  {Zhang}}]{doi:10.1126/science.1133734}%
  \BibitemOpen
  \bibfield  {author} {\bibinfo {author} {\bibfnamefont {B.~A.}\ \bibnamefont
  {Bernevig}}, \bibinfo {author} {\bibfnamefont {T.~L.}\ \bibnamefont
  {Hughes}},\ and\ \bibinfo {author} {\bibfnamefont {S.-C.}\ \bibnamefont
  {Zhang}},\ }\bibfield  {title} {\bibinfo {title} {Quantum spin hall effect
  and topological phase transition in hgte quantum wells},\ }\href
  {https://doi.org/10.1126/science.1133734} {\bibfield  {journal} {\bibinfo
  {journal} {Science}\ }\textbf {\bibinfo {volume} {314}},\ \bibinfo {pages}
  {1757} (\bibinfo {year} {2006})}\BibitemShut {NoStop}%
\bibitem [{\citenamefont {Teo}\ \emph {et~al.}(2008)\citenamefont {Teo},
  \citenamefont {Fu},\ and\ \citenamefont {Kane}}]{PhysRevB.78.045426}%
  \BibitemOpen
  \bibfield  {author} {\bibinfo {author} {\bibfnamefont {J.~C.~Y.}\
  \bibnamefont {Teo}}, \bibinfo {author} {\bibfnamefont {L.}~\bibnamefont
  {Fu}},\ and\ \bibinfo {author} {\bibfnamefont {C.~L.}\ \bibnamefont {Kane}},\
  }\bibfield  {title} {\bibinfo {title} {Surface states and topological
  invariants in three-dimensional topological insulators: Application to
  ${\text{bi}}_{1\ensuremath{-}x}{\text{sb}}_{x}$},\ }\href
  {https://doi.org/10.1103/PhysRevB.78.045426} {\bibfield  {journal} {\bibinfo
  {journal} {Phys. Rev. B}\ }\textbf {\bibinfo {volume} {78}},\ \bibinfo
  {pages} {045426} (\bibinfo {year} {2008})}\BibitemShut {NoStop}%
\bibitem [{\citenamefont {Fu}(2011)}]{PhysRevLett.106.106802}%
  \BibitemOpen
  \bibfield  {author} {\bibinfo {author} {\bibfnamefont {L.}~\bibnamefont
  {Fu}},\ }\bibfield  {title} {\bibinfo {title} {Topological crystalline
  insulators},\ }\href {https://doi.org/10.1103/PhysRevLett.106.106802}
  {\bibfield  {journal} {\bibinfo  {journal} {Phys. Rev. Lett.}\ }\textbf
  {\bibinfo {volume} {106}},\ \bibinfo {pages} {106802} (\bibinfo {year}
  {2011})}\BibitemShut {NoStop}%
\bibitem [{\citenamefont {Benalcazar}\ \emph {et~al.}(2017)\citenamefont
  {Benalcazar}, \citenamefont {Bernevig},\ and\ \citenamefont
  {Hughes}}]{PhysRevB.96.245115}%
  \BibitemOpen
  \bibfield  {author} {\bibinfo {author} {\bibfnamefont {W.~A.}\ \bibnamefont
  {Benalcazar}}, \bibinfo {author} {\bibfnamefont {B.~A.}\ \bibnamefont
  {Bernevig}},\ and\ \bibinfo {author} {\bibfnamefont {T.~L.}\ \bibnamefont
  {Hughes}},\ }\bibfield  {title} {\bibinfo {title} {Electric multipole
  moments, topological multipole moment pumping, and chiral hinge states in
  crystalline insulators},\ }\href {https://doi.org/10.1103/PhysRevB.96.245115}
  {\bibfield  {journal} {\bibinfo  {journal} {Phys. Rev. B}\ }\textbf {\bibinfo
  {volume} {96}},\ \bibinfo {pages} {245115} (\bibinfo {year}
  {2017})}\BibitemShut {NoStop}%
\bibitem [{\citenamefont {Fang}\ and\ \citenamefont
  {Fu}(2015)}]{PhysRevB.91.161105}%
  \BibitemOpen
  \bibfield  {author} {\bibinfo {author} {\bibfnamefont {C.}~\bibnamefont
  {Fang}}\ and\ \bibinfo {author} {\bibfnamefont {L.}~\bibnamefont {Fu}},\
  }\bibfield  {title} {\bibinfo {title} {New classes of three-dimensional
  topological crystalline insulators: Nonsymmorphic and magnetic},\ }\href
  {https://doi.org/10.1103/PhysRevB.91.161105} {\bibfield  {journal} {\bibinfo
  {journal} {Phys. Rev. B}\ }\textbf {\bibinfo {volume} {91}},\ \bibinfo
  {pages} {161105} (\bibinfo {year} {2015})}\BibitemShut {NoStop}%
\bibitem [{\citenamefont {Watanabe}\ and\ \citenamefont
  {Fu}(2017)}]{PhysRevB.95.081107}%
  \BibitemOpen
  \bibfield  {author} {\bibinfo {author} {\bibfnamefont {H.}~\bibnamefont
  {Watanabe}}\ and\ \bibinfo {author} {\bibfnamefont {L.}~\bibnamefont {Fu}},\
  }\bibfield  {title} {\bibinfo {title} {Topological crystalline magnets:
  Symmetry-protected topological phases of fermions},\ }\href
  {https://doi.org/10.1103/PhysRevB.95.081107} {\bibfield  {journal} {\bibinfo
  {journal} {Phys. Rev. B}\ }\textbf {\bibinfo {volume} {95}},\ \bibinfo
  {pages} {081107} (\bibinfo {year} {2017})}\BibitemShut {NoStop}%
\bibitem [{\citenamefont {Shiozaki}\ and\ \citenamefont
  {Sato}(2014)}]{PhysRevB.90.165114}%
  \BibitemOpen
  \bibfield  {author} {\bibinfo {author} {\bibfnamefont {K.}~\bibnamefont
  {Shiozaki}}\ and\ \bibinfo {author} {\bibfnamefont {M.}~\bibnamefont
  {Sato}},\ }\bibfield  {title} {\bibinfo {title} {Topology of crystalline
  insulators and superconductors},\ }\href
  {https://doi.org/10.1103/PhysRevB.90.165114} {\bibfield  {journal} {\bibinfo
  {journal} {Phys. Rev. B}\ }\textbf {\bibinfo {volume} {90}},\ \bibinfo
  {pages} {165114} (\bibinfo {year} {2014})}\BibitemShut {NoStop}%
\bibitem [{\citenamefont {van Miert}\ and\ \citenamefont
  {Ortix}(2018)}]{PhysRevB.98.081110}%
  \BibitemOpen
  \bibfield  {author} {\bibinfo {author} {\bibfnamefont {G.}~\bibnamefont {van
  Miert}}\ and\ \bibinfo {author} {\bibfnamefont {C.}~\bibnamefont {Ortix}},\
  }\bibfield  {title} {\bibinfo {title} {Higher-order topological insulators
  protected by inversion and rotoinversion symmetries},\ }\href
  {https://doi.org/10.1103/PhysRevB.98.081110} {\bibfield  {journal} {\bibinfo
  {journal} {Phys. Rev. B}\ }\textbf {\bibinfo {volume} {98}},\ \bibinfo
  {pages} {081110} (\bibinfo {year} {2018})}\BibitemShut {NoStop}%
\bibitem [{\citenamefont {Song}\ \emph {et~al.}(2017)\citenamefont {Song},
  \citenamefont {Fang},\ and\ \citenamefont {Fang}}]{PhysRevLett.119.246402}%
  \BibitemOpen
  \bibfield  {author} {\bibinfo {author} {\bibfnamefont {Z.}~\bibnamefont
  {Song}}, \bibinfo {author} {\bibfnamefont {Z.}~\bibnamefont {Fang}},\ and\
  \bibinfo {author} {\bibfnamefont {C.}~\bibnamefont {Fang}},\ }\bibfield
  {title} {\bibinfo {title} {$(d\ensuremath{-}2)$-dimensional edge states of
  rotation symmetry protected topological states},\ }\href
  {https://doi.org/10.1103/PhysRevLett.119.246402} {\bibfield  {journal}
  {\bibinfo  {journal} {Phys. Rev. Lett.}\ }\textbf {\bibinfo {volume} {119}},\
  \bibinfo {pages} {246402} (\bibinfo {year} {2017})}\BibitemShut {NoStop}%
\bibitem [{\citenamefont {Hirayama}\ \emph {et~al.}(2020)\citenamefont
  {Hirayama}, \citenamefont {Takahashi}, \citenamefont {Matsuishi},
  \citenamefont {Hosono},\ and\ \citenamefont
  {Murakami}}]{PhysRevResearch.2.043131}%
  \BibitemOpen
  \bibfield  {author} {\bibinfo {author} {\bibfnamefont {M.}~\bibnamefont
  {Hirayama}}, \bibinfo {author} {\bibfnamefont {R.}~\bibnamefont {Takahashi}},
  \bibinfo {author} {\bibfnamefont {S.}~\bibnamefont {Matsuishi}}, \bibinfo
  {author} {\bibfnamefont {H.}~\bibnamefont {Hosono}},\ and\ \bibinfo {author}
  {\bibfnamefont {S.}~\bibnamefont {Murakami}},\ }\bibfield  {title} {\bibinfo
  {title} {Higher-order topological crystalline insulating phase and quantized
  hinge charge in topological electride apatite},\ }\href
  {https://doi.org/10.1103/PhysRevResearch.2.043131} {\bibfield  {journal}
  {\bibinfo  {journal} {Phys. Rev. Res.}\ }\textbf {\bibinfo {volume} {2}},\
  \bibinfo {pages} {043131} (\bibinfo {year} {2020})}\BibitemShut {NoStop}%
\bibitem [{\citenamefont {Naito}\ \emph {et~al.}(2022)\citenamefont {Naito},
  \citenamefont {Takahashi}, \citenamefont {Watanabe},\ and\ \citenamefont
  {Murakami}}]{PhysRevB.105.045126}%
  \BibitemOpen
  \bibfield  {author} {\bibinfo {author} {\bibfnamefont {K.}~\bibnamefont
  {Naito}}, \bibinfo {author} {\bibfnamefont {R.}~\bibnamefont {Takahashi}},
  \bibinfo {author} {\bibfnamefont {H.}~\bibnamefont {Watanabe}},\ and\
  \bibinfo {author} {\bibfnamefont {S.}~\bibnamefont {Murakami}},\ }\bibfield
  {title} {\bibinfo {title} {Fractional hinge and corner charges in various
  crystal shapes with cubic symmetry},\ }\href
  {https://doi.org/10.1103/PhysRevB.105.045126} {\bibfield  {journal} {\bibinfo
   {journal} {Phys. Rev. B}\ }\textbf {\bibinfo {volume} {105}},\ \bibinfo
  {pages} {045126} (\bibinfo {year} {2022})}\BibitemShut {NoStop}%
\bibitem [{\citenamefont {Benalcazar}\ \emph {et~al.}(2019)\citenamefont
  {Benalcazar}, \citenamefont {Li},\ and\ \citenamefont
  {Hughes}}]{PhysRevB.99.245151}%
  \BibitemOpen
  \bibfield  {author} {\bibinfo {author} {\bibfnamefont {W.~A.}\ \bibnamefont
  {Benalcazar}}, \bibinfo {author} {\bibfnamefont {T.}~\bibnamefont {Li}},\
  and\ \bibinfo {author} {\bibfnamefont {T.~L.}\ \bibnamefont {Hughes}},\
  }\bibfield  {title} {\bibinfo {title} {Quantization of fractional corner
  charge in ${C}_{n}$-symmetric higher-order topological crystalline
  insulators},\ }\href {https://doi.org/10.1103/PhysRevB.99.245151} {\bibfield
  {journal} {\bibinfo  {journal} {Phys. Rev. B}\ }\textbf {\bibinfo {volume}
  {99}},\ \bibinfo {pages} {245151} (\bibinfo {year} {2019})}\BibitemShut
  {NoStop}%
\bibitem [{\citenamefont {Schindler}\ \emph
  {et~al.}(2018{\natexlab{a}})\citenamefont {Schindler}, \citenamefont {Wang},
  \citenamefont {Vergniory}, \citenamefont {Cook}, \citenamefont {Murani},
  \citenamefont {Sengupta}, \citenamefont {Kasumov}, \citenamefont {Deblock},
  \citenamefont {Jeon}, \citenamefont {Drozdov}, \citenamefont {Bouchiat},
  \citenamefont {Gu{\'{e}}ron}, \citenamefont {Yazdani}, \citenamefont
  {Bernevig},\ and\ \citenamefont {Neupert}}]{Schindler_2018}%
  \BibitemOpen
  \bibfield  {author} {\bibinfo {author} {\bibfnamefont {F.}~\bibnamefont
  {Schindler}}, \bibinfo {author} {\bibfnamefont {Z.}~\bibnamefont {Wang}},
  \bibinfo {author} {\bibfnamefont {M.~G.}\ \bibnamefont {Vergniory}}, \bibinfo
  {author} {\bibfnamefont {A.~M.}\ \bibnamefont {Cook}}, \bibinfo {author}
  {\bibfnamefont {A.}~\bibnamefont {Murani}}, \bibinfo {author} {\bibfnamefont
  {S.}~\bibnamefont {Sengupta}}, \bibinfo {author} {\bibfnamefont {A.~Y.}\
  \bibnamefont {Kasumov}}, \bibinfo {author} {\bibfnamefont {R.}~\bibnamefont
  {Deblock}}, \bibinfo {author} {\bibfnamefont {S.}~\bibnamefont {Jeon}},
  \bibinfo {author} {\bibfnamefont {I.}~\bibnamefont {Drozdov}}, \bibinfo
  {author} {\bibfnamefont {H.}~\bibnamefont {Bouchiat}}, \bibinfo {author}
  {\bibfnamefont {S.}~\bibnamefont {Gu{\'{e}}ron}}, \bibinfo {author}
  {\bibfnamefont {A.}~\bibnamefont {Yazdani}}, \bibinfo {author} {\bibfnamefont
  {B.~A.}\ \bibnamefont {Bernevig}},\ and\ \bibinfo {author} {\bibfnamefont
  {T.}~\bibnamefont {Neupert}},\ }\bibfield  {title} {\bibinfo {title}
  {Higher-order topology in bismuth},\ }\href
  {https://doi.org/10.1038/s41567-018-0224-7} {\bibfield  {journal} {\bibinfo
  {journal} {Nature Physics}\ }\textbf {\bibinfo {volume} {14}},\ \bibinfo
  {pages} {918} (\bibinfo {year} {2018}{\natexlab{a}})}\BibitemShut {NoStop}%
\bibitem [{\citenamefont {Schindler}\ \emph
  {et~al.}(2018{\natexlab{b}})\citenamefont {Schindler}, \citenamefont {Cook},
  \citenamefont {Vergniory}, \citenamefont {Wang}, \citenamefont {Parkin},
  \citenamefont {Bernevig},\ and\ \citenamefont
  {Neupert}}]{doi:10.1126/sciadv.aat0346}%
  \BibitemOpen
  \bibfield  {author} {\bibinfo {author} {\bibfnamefont {F.}~\bibnamefont
  {Schindler}}, \bibinfo {author} {\bibfnamefont {A.~M.}\ \bibnamefont {Cook}},
  \bibinfo {author} {\bibfnamefont {M.~G.}\ \bibnamefont {Vergniory}}, \bibinfo
  {author} {\bibfnamefont {Z.}~\bibnamefont {Wang}}, \bibinfo {author}
  {\bibfnamefont {S.~S.~P.}\ \bibnamefont {Parkin}}, \bibinfo {author}
  {\bibfnamefont {B.~A.}\ \bibnamefont {Bernevig}},\ and\ \bibinfo {author}
  {\bibfnamefont {T.}~\bibnamefont {Neupert}},\ }\bibfield  {title} {\bibinfo
  {title} {Higher-order topological insulators},\ }\href
  {https://doi.org/10.1126/sciadv.aat0346} {\bibfield  {journal} {\bibinfo
  {journal} {Science Advances}\ }\textbf {\bibinfo {volume} {4}},\ \bibinfo
  {pages} {eaat0346} (\bibinfo {year} {2018}{\natexlab{b}})}\BibitemShut
  {NoStop}%
\bibitem [{\citenamefont {Agarwala}\ \emph {et~al.}(2020)\citenamefont
  {Agarwala}, \citenamefont {Juri\ifmmode \check{c}\else
  \v{c}\fi{}i\ifmmode~\acute{c}\else \'{c}\fi{}},\ and\ \citenamefont
  {Roy}}]{PhysRevResearch.2.012067}%
  \BibitemOpen
  \bibfield  {author} {\bibinfo {author} {\bibfnamefont {A.}~\bibnamefont
  {Agarwala}}, \bibinfo {author} {\bibfnamefont {V.}~\bibnamefont {Juri\ifmmode
  \check{c}\else \v{c}\fi{}i\ifmmode~\acute{c}\else \'{c}\fi{}}},\ and\
  \bibinfo {author} {\bibfnamefont {B.}~\bibnamefont {Roy}},\ }\bibfield
  {title} {\bibinfo {title} {Higher-order topological insulators in amorphous
  solids},\ }\href {https://doi.org/10.1103/PhysRevResearch.2.012067}
  {\bibfield  {journal} {\bibinfo  {journal} {Phys. Rev. Res.}\ }\textbf
  {\bibinfo {volume} {2}},\ \bibinfo {pages} {012067} (\bibinfo {year}
  {2020})}\BibitemShut {NoStop}%
\bibitem [{\citenamefont {Chen}\ \emph {et~al.}(2020)\citenamefont {Chen},
  \citenamefont {Chen}, \citenamefont {Gao}, \citenamefont {Zhou},\ and\
  \citenamefont {Xu}}]{PhysRevLett.124.036803}%
  \BibitemOpen
  \bibfield  {author} {\bibinfo {author} {\bibfnamefont {R.}~\bibnamefont
  {Chen}}, \bibinfo {author} {\bibfnamefont {C.-Z.}\ \bibnamefont {Chen}},
  \bibinfo {author} {\bibfnamefont {J.-H.}\ \bibnamefont {Gao}}, \bibinfo
  {author} {\bibfnamefont {B.}~\bibnamefont {Zhou}},\ and\ \bibinfo {author}
  {\bibfnamefont {D.-H.}\ \bibnamefont {Xu}},\ }\bibfield  {title} {\bibinfo
  {title} {Higher-order topological insulators in quasicrystals},\ }\href
  {https://doi.org/10.1103/PhysRevLett.124.036803} {\bibfield  {journal}
  {\bibinfo  {journal} {Phys. Rev. Lett.}\ }\textbf {\bibinfo {volume} {124}},\
  \bibinfo {pages} {036803} (\bibinfo {year} {2020})}\BibitemShut {NoStop}%
\bibitem [{\citenamefont {Watanabe}\ and\ \citenamefont
  {Po}(2021)}]{PhysRevX.11.041064}%
  \BibitemOpen
  \bibfield  {author} {\bibinfo {author} {\bibfnamefont {H.}~\bibnamefont
  {Watanabe}}\ and\ \bibinfo {author} {\bibfnamefont {H.~C.}\ \bibnamefont
  {Po}},\ }\bibfield  {title} {\bibinfo {title} {Fractional corner charge of
  sodium chloride},\ }\href {https://doi.org/10.1103/PhysRevX.11.041064}
  {\bibfield  {journal} {\bibinfo  {journal} {Phys. Rev. X}\ }\textbf {\bibinfo
  {volume} {11}},\ \bibinfo {pages} {041064} (\bibinfo {year}
  {2021})}\BibitemShut {NoStop}%
\bibitem [{\citenamefont {Takahashi}\ \emph {et~al.}(2021)\citenamefont
  {Takahashi}, \citenamefont {Zhang},\ and\ \citenamefont
  {Murakami}}]{PhysRevB.103.205123}%
  \BibitemOpen
  \bibfield  {author} {\bibinfo {author} {\bibfnamefont {R.}~\bibnamefont
  {Takahashi}}, \bibinfo {author} {\bibfnamefont {T.}~\bibnamefont {Zhang}},\
  and\ \bibinfo {author} {\bibfnamefont {S.}~\bibnamefont {Murakami}},\
  }\bibfield  {title} {\bibinfo {title} {General corner charge formula in
  two-dimensional ${C}_{n}$-symmetric higher-order topological insulators},\
  }\href {https://doi.org/10.1103/PhysRevB.103.205123} {\bibfield  {journal}
  {\bibinfo  {journal} {Phys. Rev. B}\ }\textbf {\bibinfo {volume} {103}},\
  \bibinfo {pages} {205123} (\bibinfo {year} {2021})}\BibitemShut {NoStop}%
\bibitem [{\citenamefont {Schindler}\ \emph {et~al.}(2019)\citenamefont
  {Schindler}, \citenamefont {Brzezi\ifmmode~\acute{n}\else \'{n}\fi{}ska},
  \citenamefont {Benalcazar}, \citenamefont {Iraola}, \citenamefont {Bouhon},
  \citenamefont {Tsirkin}, \citenamefont {Vergniory},\ and\ \citenamefont
  {Neupert}}]{PhysRevResearch.1.033074}%
  \BibitemOpen
  \bibfield  {author} {\bibinfo {author} {\bibfnamefont {F.}~\bibnamefont
  {Schindler}}, \bibinfo {author} {\bibfnamefont {M.}~\bibnamefont
  {Brzezi\ifmmode~\acute{n}\else \'{n}\fi{}ska}}, \bibinfo {author}
  {\bibfnamefont {W.~A.}\ \bibnamefont {Benalcazar}}, \bibinfo {author}
  {\bibfnamefont {M.}~\bibnamefont {Iraola}}, \bibinfo {author} {\bibfnamefont
  {A.}~\bibnamefont {Bouhon}}, \bibinfo {author} {\bibfnamefont {S.~S.}\
  \bibnamefont {Tsirkin}}, \bibinfo {author} {\bibfnamefont {M.~G.}\
  \bibnamefont {Vergniory}},\ and\ \bibinfo {author} {\bibfnamefont
  {T.}~\bibnamefont {Neupert}},\ }\bibfield  {title} {\bibinfo {title}
  {Fractional corner charges in spin-orbit coupled crystals},\ }\href
  {https://doi.org/10.1103/PhysRevResearch.1.033074} {\bibfield  {journal}
  {\bibinfo  {journal} {Phys. Rev. Res.}\ }\textbf {\bibinfo {volume} {1}},\
  \bibinfo {pages} {033074} (\bibinfo {year} {2019})}\BibitemShut {NoStop}%
\bibitem [{\citenamefont {Watanabe}\ and\ \citenamefont
  {Ono}(2020)}]{PhysRevB.102.165120}%
  \BibitemOpen
  \bibfield  {author} {\bibinfo {author} {\bibfnamefont {H.}~\bibnamefont
  {Watanabe}}\ and\ \bibinfo {author} {\bibfnamefont {S.}~\bibnamefont {Ono}},\
  }\bibfield  {title} {\bibinfo {title} {Corner charge and bulk multipole
  moment in periodic systems},\ }\href
  {https://doi.org/10.1103/PhysRevB.102.165120} {\bibfield  {journal} {\bibinfo
   {journal} {Phys. Rev. B}\ }\textbf {\bibinfo {volume} {102}},\ \bibinfo
  {pages} {165120} (\bibinfo {year} {2020})}\BibitemShut {NoStop}%
\bibitem [{\citenamefont
  {Volovik}(2009)}]{10.1093/acprof:oso/9780199564842.001.0001}%
  \BibitemOpen
  \bibfield  {author} {\bibinfo {author} {\bibfnamefont {G.~E.}\ \bibnamefont
  {Volovik}},\ }\href
  {https://doi.org/10.1093/acprof:oso/9780199564842.001.0001} {\emph {\bibinfo
  {title} {{The Universe in a Helium Droplet}}}}\ (\bibinfo  {publisher}
  {Oxford University Press},\ \bibinfo {year} {2009})\BibitemShut {NoStop}%
\bibitem [{\citenamefont {Schnyder}\ \emph {et~al.}(2008)\citenamefont
  {Schnyder}, \citenamefont {Ryu}, \citenamefont {Furusaki},\ and\
  \citenamefont {Ludwig}}]{PhysRevB.78.195125}%
  \BibitemOpen
  \bibfield  {author} {\bibinfo {author} {\bibfnamefont {A.~P.}\ \bibnamefont
  {Schnyder}}, \bibinfo {author} {\bibfnamefont {S.}~\bibnamefont {Ryu}},
  \bibinfo {author} {\bibfnamefont {A.}~\bibnamefont {Furusaki}},\ and\
  \bibinfo {author} {\bibfnamefont {A.~W.~W.}\ \bibnamefont {Ludwig}},\
  }\bibfield  {title} {\bibinfo {title} {Classification of topological
  insulators and superconductors in three spatial dimensions},\ }\href
  {https://doi.org/10.1103/PhysRevB.78.195125} {\bibfield  {journal} {\bibinfo
  {journal} {Phys. Rev. B}\ }\textbf {\bibinfo {volume} {78}},\ \bibinfo
  {pages} {195125} (\bibinfo {year} {2008})}\BibitemShut {NoStop}%
\bibitem [{\citenamefont {Yokomizo}\ and\ \citenamefont
  {Murakami}(2019)}]{PhysRevLett.123.066404}%
  \BibitemOpen
  \bibfield  {author} {\bibinfo {author} {\bibfnamefont {K.}~\bibnamefont
  {Yokomizo}}\ and\ \bibinfo {author} {\bibfnamefont {S.}~\bibnamefont
  {Murakami}},\ }\bibfield  {title} {\bibinfo {title} {Non-bloch band theory of
  non-hermitian systems},\ }\href
  {https://doi.org/10.1103/PhysRevLett.123.066404} {\bibfield  {journal}
  {\bibinfo  {journal} {Phys. Rev. Lett.}\ }\textbf {\bibinfo {volume} {123}},\
  \bibinfo {pages} {066404} (\bibinfo {year} {2019})}\BibitemShut {NoStop}%
\bibitem [{\citenamefont {Zhang}\ \emph {et~al.}(2022)\citenamefont {Zhang},
  \citenamefont {Yang},\ and\ \citenamefont {Fang}}]{Zhang_2022}%
  \BibitemOpen
  \bibfield  {author} {\bibinfo {author} {\bibfnamefont {K.}~\bibnamefont
  {Zhang}}, \bibinfo {author} {\bibfnamefont {Z.}~\bibnamefont {Yang}},\ and\
  \bibinfo {author} {\bibfnamefont {C.}~\bibnamefont {Fang}},\ }\bibfield
  {title} {\bibinfo {title} {Universal non-hermitian skin effect in two and
  higher dimensions},\ }\bibfield  {journal} {\bibinfo  {journal} {Nature
  Communications}\ }\textbf {\bibinfo {volume} {13}},\ \href
  {https://doi.org/10.1038/s41467-022-30161-6} {10.1038/s41467-022-30161-6}
  (\bibinfo {year} {2022})\BibitemShut {NoStop}%
\bibitem [{\citenamefont {Yokomizo}\ and\ \citenamefont
  {Murakami}(2023)}]{PhysRevB.107.195112}%
  \BibitemOpen
  \bibfield  {author} {\bibinfo {author} {\bibfnamefont {K.}~\bibnamefont
  {Yokomizo}}\ and\ \bibinfo {author} {\bibfnamefont {S.}~\bibnamefont
  {Murakami}},\ }\bibfield  {title} {\bibinfo {title} {Non-bloch bands in
  two-dimensional non-hermitian systems},\ }\href
  {https://doi.org/10.1103/PhysRevB.107.195112} {\bibfield  {journal} {\bibinfo
   {journal} {Phys. Rev. B}\ }\textbf {\bibinfo {volume} {107}},\ \bibinfo
  {pages} {195112} (\bibinfo {year} {2023})}\BibitemShut {NoStop}%
\bibitem [{\citenamefont {Bradlyn}\ \emph {et~al.}(2017)\citenamefont
  {Bradlyn}, \citenamefont {Elcoro}, \citenamefont {Cano}, \citenamefont
  {Vergniory}, \citenamefont {Wang}, \citenamefont {Felser}, \citenamefont
  {Aroyo},\ and\ \citenamefont {Bernevig}}]{Bradlyn_2017}%
  \BibitemOpen
  \bibfield  {author} {\bibinfo {author} {\bibfnamefont {B.}~\bibnamefont
  {Bradlyn}}, \bibinfo {author} {\bibfnamefont {L.}~\bibnamefont {Elcoro}},
  \bibinfo {author} {\bibfnamefont {J.}~\bibnamefont {Cano}}, \bibinfo {author}
  {\bibfnamefont {M.~G.}\ \bibnamefont {Vergniory}}, \bibinfo {author}
  {\bibfnamefont {Z.}~\bibnamefont {Wang}}, \bibinfo {author} {\bibfnamefont
  {C.}~\bibnamefont {Felser}}, \bibinfo {author} {\bibfnamefont {M.~I.}\
  \bibnamefont {Aroyo}},\ and\ \bibinfo {author} {\bibfnamefont {B.~A.}\
  \bibnamefont {Bernevig}},\ }\bibfield  {title} {\bibinfo {title} {Topological
  quantum chemistry},\ }\href {https://doi.org/10.1038/nature23268} {\bibfield
  {journal} {\bibinfo  {journal} {Nature}\ }\textbf {\bibinfo {volume} {547}},\
  \bibinfo {pages} {298} (\bibinfo {year} {2017})}\BibitemShut {NoStop}%
\bibitem [{\citenamefont {Cano}\ \emph {et~al.}(2018)\citenamefont {Cano},
  \citenamefont {Bradlyn}, \citenamefont {Wang}, \citenamefont {Elcoro},
  \citenamefont {Vergniory}, \citenamefont {Felser}, \citenamefont {Aroyo},\
  and\ \citenamefont {Bernevig}}]{PhysRevB.97.035139}%
  \BibitemOpen
  \bibfield  {author} {\bibinfo {author} {\bibfnamefont {J.}~\bibnamefont
  {Cano}}, \bibinfo {author} {\bibfnamefont {B.}~\bibnamefont {Bradlyn}},
  \bibinfo {author} {\bibfnamefont {Z.}~\bibnamefont {Wang}}, \bibinfo {author}
  {\bibfnamefont {L.}~\bibnamefont {Elcoro}}, \bibinfo {author} {\bibfnamefont
  {M.~G.}\ \bibnamefont {Vergniory}}, \bibinfo {author} {\bibfnamefont
  {C.}~\bibnamefont {Felser}}, \bibinfo {author} {\bibfnamefont {M.~I.}\
  \bibnamefont {Aroyo}},\ and\ \bibinfo {author} {\bibfnamefont {B.~A.}\
  \bibnamefont {Bernevig}},\ }\bibfield  {title} {\bibinfo {title} {Building
  blocks of topological quantum chemistry: Elementary band representations},\
  }\href {https://doi.org/10.1103/PhysRevB.97.035139} {\bibfield  {journal}
  {\bibinfo  {journal} {Phys. Rev. B}\ }\textbf {\bibinfo {volume} {97}},\
  \bibinfo {pages} {035139} (\bibinfo {year} {2018})}\BibitemShut {NoStop}%
\bibitem [{\citenamefont {Aroyo}\ \emph {et~al.}(2006)\citenamefont {Aroyo},
  \citenamefont {Kirov}, \citenamefont {Capillas}, \citenamefont {Perez-Mato},\
  and\ \citenamefont {Wondratschek}}]{Aroyo:xo5013}%
  \BibitemOpen
  \bibfield  {author} {\bibinfo {author} {\bibfnamefont {M.~I.}\ \bibnamefont
  {Aroyo}}, \bibinfo {author} {\bibfnamefont {A.}~\bibnamefont {Kirov}},
  \bibinfo {author} {\bibfnamefont {C.}~\bibnamefont {Capillas}}, \bibinfo
  {author} {\bibfnamefont {J.~M.}\ \bibnamefont {Perez-Mato}},\ and\ \bibinfo
  {author} {\bibfnamefont {H.}~\bibnamefont {Wondratschek}},\ }\bibfield
  {title} {\bibinfo {title} {{Bilbao Crystallographic Server. II.
  Representations of crystallographic point groups and space groups}},\ }\href
  {https://doi.org/10.1107/S0108767305040286} {\bibfield  {journal} {\bibinfo
  {journal} {Acta Crystallographica Section A}\ }\textbf {\bibinfo {volume}
  {62}},\ \bibinfo {pages} {115} (\bibinfo {year} {2006})}\BibitemShut
  {NoStop}%
\bibitem [{\citenamefont {Po}\ \emph {et~al.}(2017)\citenamefont {Po},
  \citenamefont {Vishwanath},\ and\ \citenamefont {Watanabe}}]{Po_2017}%
  \BibitemOpen
  \bibfield  {author} {\bibinfo {author} {\bibfnamefont {H.~C.}\ \bibnamefont
  {Po}}, \bibinfo {author} {\bibfnamefont {A.}~\bibnamefont {Vishwanath}},\
  and\ \bibinfo {author} {\bibfnamefont {H.}~\bibnamefont {Watanabe}},\
  }\bibfield  {title} {\bibinfo {title} {Symmetry-based indicators of band
  topology in the 230 space groups},\ }\bibfield  {journal} {\bibinfo
  {journal} {Nature Communications}\ }\textbf {\bibinfo {volume} {8}},\ \href
  {https://doi.org/10.1038/s41467-017-00133-2} {10.1038/s41467-017-00133-2}
  (\bibinfo {year} {2017})\BibitemShut {NoStop}%
\bibitem [{\citenamefont {Slager}\ \emph {et~al.}(2013)\citenamefont {Slager},
  \citenamefont {Mesaros}, \citenamefont {Juri{\v c}i{\'c}},\ and\
  \citenamefont {Zaanen}}]{Slager:2013aa}%
  \BibitemOpen
  \bibfield  {author} {\bibinfo {author} {\bibfnamefont {R.-J.}\ \bibnamefont
  {Slager}}, \bibinfo {author} {\bibfnamefont {A.}~\bibnamefont {Mesaros}},
  \bibinfo {author} {\bibfnamefont {V.}~\bibnamefont {Juri{\v c}i{\'c}}},\ and\
  \bibinfo {author} {\bibfnamefont {J.}~\bibnamefont {Zaanen}},\ }\bibfield
  {title} {\bibinfo {title} {The space group classification of topological
  band-insulators},\ }\href {https://doi.org/10.1038/nphys2513} {\bibfield
  {journal} {\bibinfo  {journal} {Nature Physics}\ }\textbf {\bibinfo {volume}
  {9}},\ \bibinfo {pages} {98} (\bibinfo {year} {2013})}\BibitemShut {NoStop}%
\bibitem [{\citenamefont {Kruthoff}\ \emph {et~al.}(2017)\citenamefont
  {Kruthoff}, \citenamefont {de~Boer}, \citenamefont {van Wezel}, \citenamefont
  {Kane},\ and\ \citenamefont {Slager}}]{PhysRevX.7.041069}%
  \BibitemOpen
  \bibfield  {author} {\bibinfo {author} {\bibfnamefont {J.}~\bibnamefont
  {Kruthoff}}, \bibinfo {author} {\bibfnamefont {J.}~\bibnamefont {de~Boer}},
  \bibinfo {author} {\bibfnamefont {J.}~\bibnamefont {van Wezel}}, \bibinfo
  {author} {\bibfnamefont {C.~L.}\ \bibnamefont {Kane}},\ and\ \bibinfo
  {author} {\bibfnamefont {R.-J.}\ \bibnamefont {Slager}},\ }\bibfield  {title}
  {\bibinfo {title} {{Topological Classification of Crystalline Insulators
  through Band Structure Combinatorics}},\ }\href
  {https://doi.org/10.1103/PhysRevX.7.041069} {\bibfield  {journal} {\bibinfo
  {journal} {Phys. Rev. X}\ }\textbf {\bibinfo {volume} {7}},\ \bibinfo {pages}
  {041069} (\bibinfo {year} {2017})}\BibitemShut {NoStop}%
\bibitem [{\citenamefont {Elcoro}\ \emph {et~al.}(2021)\citenamefont {Elcoro},
  \citenamefont {Wieder}, \citenamefont {Song}, \citenamefont {Xu},
  \citenamefont {Bradlyn},\ and\ \citenamefont {Bernevig}}]{MTQC}%
  \BibitemOpen
  \bibfield  {author} {\bibinfo {author} {\bibfnamefont {L.}~\bibnamefont
  {Elcoro}}, \bibinfo {author} {\bibfnamefont {B.~J.}\ \bibnamefont {Wieder}},
  \bibinfo {author} {\bibfnamefont {Z.}~\bibnamefont {Song}}, \bibinfo {author}
  {\bibfnamefont {Y.}~\bibnamefont {Xu}}, \bibinfo {author} {\bibfnamefont
  {B.}~\bibnamefont {Bradlyn}},\ and\ \bibinfo {author} {\bibfnamefont {B.~A.}\
  \bibnamefont {Bernevig}},\ }\bibfield  {title} {\bibinfo {title} {Magnetic
  topological quantum chemistry},\ }\href
  {https://doi.org/10.1038/s41467-021-26241-8} {\bibfield  {journal} {\bibinfo
  {journal} {Nature Communications}\ }\textbf {\bibinfo {volume} {12}},\
  \bibinfo {pages} {5965} (\bibinfo {year} {2021})}\BibitemShut {NoStop}%
\bibitem [{\citenamefont {Watanabe}\ \emph {et~al.}(2018)\citenamefont
  {Watanabe}, \citenamefont {Po},\ and\ \citenamefont
  {Vishwanath}}]{SI_SA_Watanabe}%
  \BibitemOpen
  \bibfield  {author} {\bibinfo {author} {\bibfnamefont {H.}~\bibnamefont
  {Watanabe}}, \bibinfo {author} {\bibfnamefont {H.~C.}\ \bibnamefont {Po}},\
  and\ \bibinfo {author} {\bibfnamefont {A.}~\bibnamefont {Vishwanath}},\
  }\bibfield  {title} {\bibinfo {title} {{Structure and topology of band
  structures in the 1651 magnetic space groups}},\ }\href
  {http://advances.sciencemag.org/content/4/8/eaat8685} {\bibfield  {journal}
  {\bibinfo  {journal} {Sci. Adv.}\ }\textbf {\bibinfo {volume} {4}},\ \bibinfo
  {pages} {eaat8685} (\bibinfo {year} {2018})}\BibitemShut {NoStop}%
\bibitem [{\citenamefont {Tang}\ \emph
  {et~al.}(2019{\natexlab{a}})\citenamefont {Tang}, \citenamefont {Po},
  \citenamefont {Vishwanath},\ and\ \citenamefont {Wan}}]{catalogue0}%
  \BibitemOpen
  \bibfield  {author} {\bibinfo {author} {\bibfnamefont {F.}~\bibnamefont
  {Tang}}, \bibinfo {author} {\bibfnamefont {H.~C.}\ \bibnamefont {Po}},
  \bibinfo {author} {\bibfnamefont {A.}~\bibnamefont {Vishwanath}},\ and\
  \bibinfo {author} {\bibfnamefont {X.}~\bibnamefont {Wan}},\ }\bibfield
  {title} {\bibinfo {title} {{Efficient topological materials discovery using
  symmetry indicators}},\ }\href {https://doi.org/10.1038/s41567-019-0418-7}
  {\bibfield  {journal} {\bibinfo  {journal} {Nat. Phys.}\ }\textbf {\bibinfo
  {volume} {15}},\ \bibinfo {pages} {470} (\bibinfo {year}
  {2019}{\natexlab{a}})}\BibitemShut {NoStop}%
\bibitem [{\citenamefont {Zhang}\ \emph {et~al.}(2019)\citenamefont {Zhang},
  \citenamefont {Jiang}, \citenamefont {Song}, \citenamefont {Huang},
  \citenamefont {He}, \citenamefont {Fang}, \citenamefont {Weng},\ and\
  \citenamefont {Fang}}]{catalogue1}%
  \BibitemOpen
  \bibfield  {author} {\bibinfo {author} {\bibfnamefont {T.}~\bibnamefont
  {Zhang}}, \bibinfo {author} {\bibfnamefont {Y.}~\bibnamefont {Jiang}},
  \bibinfo {author} {\bibfnamefont {Z.}~\bibnamefont {Song}}, \bibinfo {author}
  {\bibfnamefont {H.}~\bibnamefont {Huang}}, \bibinfo {author} {\bibfnamefont
  {Y.}~\bibnamefont {He}}, \bibinfo {author} {\bibfnamefont {Z.}~\bibnamefont
  {Fang}}, \bibinfo {author} {\bibfnamefont {H.}~\bibnamefont {Weng}},\ and\
  \bibinfo {author} {\bibfnamefont {C.}~\bibnamefont {Fang}},\ }\bibfield
  {title} {\bibinfo {title} {{Catalogue of topological electronic materials}},\
  }\href {https://doi.org/10.1038/s41586-019-0944-6} {\bibfield  {journal}
  {\bibinfo  {journal} {Nature}\ }\textbf {\bibinfo {volume} {566}},\ \bibinfo
  {pages} {475} (\bibinfo {year} {2019})}\BibitemShut {NoStop}%
\bibitem [{\citenamefont {Tang}\ \emph
  {et~al.}(2019{\natexlab{b}})\citenamefont {Tang}, \citenamefont {Po},
  \citenamefont {Vishwanath},\ and\ \citenamefont {Wan}}]{catalogue2}%
  \BibitemOpen
  \bibfield  {author} {\bibinfo {author} {\bibfnamefont {F.}~\bibnamefont
  {Tang}}, \bibinfo {author} {\bibfnamefont {H.~C.}\ \bibnamefont {Po}},
  \bibinfo {author} {\bibfnamefont {A.}~\bibnamefont {Vishwanath}},\ and\
  \bibinfo {author} {\bibfnamefont {X.}~\bibnamefont {Wan}},\ }\bibfield
  {title} {\bibinfo {title} {{Comprehensive search for topological materials
  using symmetry indicators}},\ }\href
  {https://doi.org/10.1038/s41586-019-0937-5} {\bibfield  {journal} {\bibinfo
  {journal} {Nature}\ }\textbf {\bibinfo {volume} {566}},\ \bibinfo {pages}
  {486} (\bibinfo {year} {2019}{\natexlab{b}})}\BibitemShut {NoStop}%
\bibitem [{\citenamefont {Vergniory}\ \emph {et~al.}(2019)\citenamefont
  {Vergniory}, \citenamefont {Elcoro}, \citenamefont {Felser}, \citenamefont
  {Regnault}, \citenamefont {Bernevig},\ and\ \citenamefont
  {Wang}}]{catalogue3}%
  \BibitemOpen
  \bibfield  {author} {\bibinfo {author} {\bibfnamefont {M.~G.}\ \bibnamefont
  {Vergniory}}, \bibinfo {author} {\bibfnamefont {L.}~\bibnamefont {Elcoro}},
  \bibinfo {author} {\bibfnamefont {C.}~\bibnamefont {Felser}}, \bibinfo
  {author} {\bibfnamefont {N.}~\bibnamefont {Regnault}}, \bibinfo {author}
  {\bibfnamefont {B.~A.}\ \bibnamefont {Bernevig}},\ and\ \bibinfo {author}
  {\bibfnamefont {Z.}~\bibnamefont {Wang}},\ }\bibfield  {title} {\bibinfo
  {title} {{A complete catalogue of high-quality topological materials}},\
  }\href {https://doi.org/10.1038/s41586-019-0954-4} {\bibfield  {journal}
  {\bibinfo  {journal} {Nature}\ }\textbf {\bibinfo {volume} {566}},\ \bibinfo
  {pages} {480} (\bibinfo {year} {2019})}\BibitemShut {NoStop}%
\bibitem [{\citenamefont {Tang}\ \emph
  {et~al.}(2019{\natexlab{c}})\citenamefont {Tang}, \citenamefont {Po},
  \citenamefont {Vishwanath},\ and\ \citenamefont {Wan}}]{catalogue4}%
  \BibitemOpen
  \bibfield  {author} {\bibinfo {author} {\bibfnamefont {F.}~\bibnamefont
  {Tang}}, \bibinfo {author} {\bibfnamefont {H.~C.}\ \bibnamefont {Po}},
  \bibinfo {author} {\bibfnamefont {A.}~\bibnamefont {Vishwanath}},\ and\
  \bibinfo {author} {\bibfnamefont {X.}~\bibnamefont {Wan}},\ }\bibfield
  {title} {\bibinfo {title} {{Topological materials discovery by large-order
  symmetry indicators}},\ }\href
  {https://advances.sciencemag.org/content/5/3/eaau8725} {\bibfield  {journal}
  {\bibinfo  {journal} {Sci. Adv.}\ }\textbf {\bibinfo {volume} {5}},\ \bibinfo
  {pages} {eaau8725} (\bibinfo {year} {2019}{\natexlab{c}})}\BibitemShut
  {NoStop}%
\bibitem [{\citenamefont {Xu}\ \emph {et~al.}(2020)\citenamefont {Xu},
  \citenamefont {Elcoro}, \citenamefont {Song}, \citenamefont {Wieder},
  \citenamefont {Vergniory}, \citenamefont {Regnault}, \citenamefont {Chen},
  \citenamefont {Felser},\ and\ \citenamefont {Bernevig}}]{catalogue6}%
  \BibitemOpen
  \bibfield  {author} {\bibinfo {author} {\bibfnamefont {Y.}~\bibnamefont
  {Xu}}, \bibinfo {author} {\bibfnamefont {L.}~\bibnamefont {Elcoro}}, \bibinfo
  {author} {\bibfnamefont {Z.-D.}\ \bibnamefont {Song}}, \bibinfo {author}
  {\bibfnamefont {B.~J.}\ \bibnamefont {Wieder}}, \bibinfo {author}
  {\bibfnamefont {M.~G.}\ \bibnamefont {Vergniory}}, \bibinfo {author}
  {\bibfnamefont {N.}~\bibnamefont {Regnault}}, \bibinfo {author}
  {\bibfnamefont {Y.}~\bibnamefont {Chen}}, \bibinfo {author} {\bibfnamefont
  {C.}~\bibnamefont {Felser}},\ and\ \bibinfo {author} {\bibfnamefont {B.~A.}\
  \bibnamefont {Bernevig}},\ }\bibfield  {title} {\bibinfo {title}
  {High-throughput calculations of magnetic topological materials},\ }\href
  {https://doi.org/10.1038/s41586-020-2837-0} {\bibfield  {journal} {\bibinfo
  {journal} {Nature}\ }\textbf {\bibinfo {volume} {586}},\ \bibinfo {pages}
  {702} (\bibinfo {year} {2020})}\BibitemShut {NoStop}%
\bibitem [{\citenamefont {Fang}\ and\ \citenamefont
  {Cano}(2021)}]{PhysRevB.103.165109}%
  \BibitemOpen
  \bibfield  {author} {\bibinfo {author} {\bibfnamefont {Y.}~\bibnamefont
  {Fang}}\ and\ \bibinfo {author} {\bibfnamefont {J.}~\bibnamefont {Cano}},\
  }\bibfield  {title} {\bibinfo {title} {Filling anomaly for general two- and
  three-dimensional ${C}_{4}$ symmetric lattices},\ }\href
  {https://doi.org/10.1103/PhysRevB.103.165109} {\bibfield  {journal} {\bibinfo
   {journal} {Phys. Rev. B}\ }\textbf {\bibinfo {volume} {103}},\ \bibinfo
  {pages} {165109} (\bibinfo {year} {2021})}\BibitemShut {NoStop}%
\bibitem [{\citenamefont {Ono}\ and\ \citenamefont
  {Shiozaki}()}]{Ono=Shiozaki_top_inv}%
  \BibitemOpen
  \bibfield  {author} {\bibinfo {author} {\bibfnamefont {S.}~\bibnamefont
  {Ono}}\ and\ \bibinfo {author} {\bibfnamefont {K.}~\bibnamefont {Shiozaki}},\
  }\href@noop {} {}\bibinfo {note} {Unpublished}\BibitemShut {NoStop}%
\bibitem [{\citenamefont {Po}\ \emph {et~al.}(2018)\citenamefont {Po},
  \citenamefont {Watanabe},\ and\ \citenamefont
  {Vishwanath}}]{FragileTI_Po_Watanabe_Vishwanath}%
  \BibitemOpen
  \bibfield  {author} {\bibinfo {author} {\bibfnamefont {H.~C.}\ \bibnamefont
  {Po}}, \bibinfo {author} {\bibfnamefont {H.}~\bibnamefont {Watanabe}},\ and\
  \bibinfo {author} {\bibfnamefont {A.}~\bibnamefont {Vishwanath}},\ }\bibfield
   {title} {\bibinfo {title} {Fragile topology and wannier obstructions},\
  }\href {https://doi.org/10.1103/PhysRevLett.121.126402} {\bibfield  {journal}
  {\bibinfo  {journal} {Phys. Rev. Lett.}\ }\textbf {\bibinfo {volume} {121}},\
  \bibinfo {pages} {126402} (\bibinfo {year} {2018})}\BibitemShut {NoStop}%
\bibitem [{\citenamefont {Vanderbilt}\ and\ \citenamefont
  {King-Smith}(1993)}]{PhysRevB.48.4442}%
  \BibitemOpen
  \bibfield  {author} {\bibinfo {author} {\bibfnamefont {D.}~\bibnamefont
  {Vanderbilt}}\ and\ \bibinfo {author} {\bibfnamefont {R.~D.}\ \bibnamefont
  {King-Smith}},\ }\bibfield  {title} {\bibinfo {title} {Electric polarization
  as a bulk quantity and its relation to surface charge},\ }\href
  {https://doi.org/10.1103/PhysRevB.48.4442} {\bibfield  {journal} {\bibinfo
  {journal} {Phys. Rev. B}\ }\textbf {\bibinfo {volume} {48}},\ \bibinfo
  {pages} {4442} (\bibinfo {year} {1993})}\BibitemShut {NoStop}%
\bibitem [{\citenamefont {King-Smith}\ and\ \citenamefont
  {Vanderbilt}(1993)}]{PhysRevB.47.1651}%
  \BibitemOpen
  \bibfield  {author} {\bibinfo {author} {\bibfnamefont {R.~D.}\ \bibnamefont
  {King-Smith}}\ and\ \bibinfo {author} {\bibfnamefont {D.}~\bibnamefont
  {Vanderbilt}},\ }\bibfield  {title} {\bibinfo {title} {Theory of polarization
  of crystalline solids},\ }\href {https://doi.org/10.1103/PhysRevB.47.1651}
  {\bibfield  {journal} {\bibinfo  {journal} {Phys. Rev. B}\ }\textbf {\bibinfo
  {volume} {47}},\ \bibinfo {pages} {1651} (\bibinfo {year}
  {1993})}\BibitemShut {NoStop}%
\bibitem [{\citenamefont {Robertson}\ \emph {et~al.}(1970)\citenamefont
  {Robertson}, \citenamefont {Carter},\ and\ \citenamefont
  {Morton}}]{ROBERTSON197079}%
  \BibitemOpen
  \bibfield  {author} {\bibinfo {author} {\bibfnamefont {S.}~\bibnamefont
  {Robertson}}, \bibinfo {author} {\bibfnamefont {S.}~\bibnamefont {Carter}},\
  and\ \bibinfo {author} {\bibfnamefont {H.}~\bibnamefont {Morton}},\
  }\bibfield  {title} {\bibinfo {title} {Finite orthogonal symmetry},\ }\href
  {https://doi.org/https://doi.org/10.1016/0040-9383(70)90052-2} {\bibfield
  {journal} {\bibinfo  {journal} {Topology}\ }\textbf {\bibinfo {volume} {9}},\
  \bibinfo {pages} {79} (\bibinfo {year} {1970})}\BibitemShut {NoStop}%
\bibitem [{\citenamefont {Robertson}\ and\ \citenamefont
  {Carter}(1970)}]{https://doi.org/10.1112/jlms/s2-2.1.125}%
  \BibitemOpen
  \bibfield  {author} {\bibinfo {author} {\bibfnamefont {S.~A.}\ \bibnamefont
  {Robertson}}\ and\ \bibinfo {author} {\bibfnamefont {S.}~\bibnamefont
  {Carter}},\ }\bibfield  {title} {\bibinfo {title} {On the platonic and
  archimedean solids},\ }\href
  {https://doi.org/https://doi.org/10.1112/jlms/s2-2.1.125} {\bibfield
  {journal} {\bibinfo  {journal} {Journal of the London Mathematical Society}\
  }\textbf {\bibinfo {volume} {s2-2}},\ \bibinfo {pages} {125} (\bibinfo {year}
  {1970})}\BibitemShut {NoStop}%
\bibitem [{\citenamefont {Bulatov}(2002)}]{bridges2002:320}%
  \BibitemOpen
  \bibfield  {author} {\bibinfo {author} {\bibfnamefont {V.}~\bibnamefont
  {Bulatov}},\ }\bibfield  {title} {\bibinfo {title} {About enumeration of
  isogonal polyhedral families - abstract},\ }in\ \href
  {http://archive.bridgesmathart.org/2002/bridges2002-320.html} {\emph
  {\bibinfo {booktitle} {Bridges: Mathematical Connections in Art, Music, and
  Science}}},\ \bibinfo {editor} {edited by\ \bibinfo {editor} {\bibfnamefont
  {R.}~\bibnamefont {Sarhangi}}}\ (\bibinfo  {publisher} {Bridges Conference},\
  \bibinfo {address} {Southwestern College, Winfield, Kansas},\ \bibinfo {year}
  {2002})\ pp.\ \bibinfo {pages} {320--320}\BibitemShut {NoStop}%
\bibitem [{\citenamefont {Cromwell}(1997)}]{polyhedra}%
  \BibitemOpen
  \bibfield  {author} {\bibinfo {author} {\bibfnamefont {P.~R.}\ \bibnamefont
  {Cromwell}},\ }\href@noop {} {\emph {\bibinfo {title} {Polyhedra}}}\
  (\bibinfo  {publisher} {Cambridge University Press, Cambridge},\ \bibinfo
  {year} {1997})\BibitemShut {NoStop}%
\bibitem [{\citenamefont {Gr{\"u}nbaum}\ and\ \citenamefont
  {Shephard}(1984)}]{Grunbaum}%
  \BibitemOpen
  \bibfield  {author} {\bibinfo {author} {\bibfnamefont {B.}~\bibnamefont
  {Gr{\"u}nbaum}}\ and\ \bibinfo {author} {\bibfnamefont {G.}~\bibnamefont
  {Shephard}},\ }\bibfield  {title} {\bibinfo {title} {Polyhedra with
  transitivity properties},\ }\href@noop {} {\bibfield  {journal} {\bibinfo
  {journal} {C. R. Math. Rep. Acad. Sci. Canada}\ }\textbf {\bibinfo {volume}
  {6}},\ \bibinfo {pages} {61–66} (\bibinfo {year} {1984})}\BibitemShut
  {NoStop}%
\bibitem [{\citenamefont {Leopold}(2017)}]{Leopold}%
  \BibitemOpen
  \bibfield  {author} {\bibinfo {author} {\bibfnamefont {U.}~\bibnamefont
  {Leopold}},\ }\bibfield  {title} {\bibinfo {title} {Vertex-transitive
  polyhedra of higher genus, i},\ }\href
  {https://doi.org/10.1007/s00454-016-9828-9} {\bibfield  {journal} {\bibinfo
  {journal} {Discrete Comput. Geom.}\ }\textbf {\bibinfo {volume} {57}},\
  \bibinfo {pages} {125} (\bibinfo {year} {2017})}\BibitemShut {NoStop}%
\bibitem [{\citenamefont {Shiozaki}\ \emph {et~al.}(2023)\citenamefont
  {Shiozaki}, \citenamefont {Xiong},\ and\ \citenamefont
  {Gomi}}]{10.1093/ptep/ptad086}%
  \BibitemOpen
  \bibfield  {author} {\bibinfo {author} {\bibfnamefont {K.}~\bibnamefont
  {Shiozaki}}, \bibinfo {author} {\bibfnamefont {C.~Z.}\ \bibnamefont
  {Xiong}},\ and\ \bibinfo {author} {\bibfnamefont {K.}~\bibnamefont {Gomi}},\
  }\bibfield  {title} {\bibinfo {title} {{Generalized homology and
  Atiyah–Hirzebruch spectral sequence in crystalline symmetry protected
  topological phenomena}},\ }\href {https://doi.org/10.1093/ptep/ptad086}
  {\bibfield  {journal} {\bibinfo  {journal} {Progress of Theoretical and
  Experimental Physics}\ }\textbf {\bibinfo {volume} {2023}},\ \bibinfo {pages}
  {083I01} (\bibinfo {year} {2023})}\BibitemShut {NoStop}%
\end{thebibliography}%

\end{document}